%% file: Paper.tex
\documentclass[twocolumn, prd,
aps,superscriptaddress,preprintnumbers,tightenlines,showpacs,nofootinbib,
eqsecnum,amsfonts,amsmath,floatfix]{revtex4}

\usepackage{epsfig}
\usepackage{graphics}
\usepackage{graphicx}
\usepackage{amssymb,mathrsfs}
\usepackage{bm}
\usepackage{xcolor,xspace}
\usepackage{wasysym}
\usepackage{times}
\usepackage{mathptmx}
\usepackage{appendix}
\usepackage{mathrsfs}
\usepackage{lipsum}
\usepackage[nolist,nohyperlinks]{acronym}

\begin{document}

\newcommand{\be}{\begin{equation}}
\newcommand{\ee}{\end{equation}}
\newcommand{\ber}{\begin{eqnarray}}
\newcommand{\eer}{\end{eqnarray}}
\newcommand{\bea}{\begin{eqnarray}}
\newcommand{\eea}{\end{eqnarray}}
\newcommand{\ie}{i.e.}
\newcommand{\dt}{{\rm d}t}
\newcommand{\df}{{\rm d}f}
\newcommand{\dtheta}{{\rm d}\theta}
\newcommand{\dphi}{{\rm d}\phi}
\newcommand{\rhat}{\hat{r}}
\newcommand{\iotahat}{\hat{\iota}}
\newcommand{\phihat}{\hat{\phi}}
\newcommand{\hc}{{\sf h}}
\newcommand{\etal}{\textit{et al.}}
\newcommand{\balpha}{{\bm \alpha}}
\newcommand{\bbeta}{{\bm \psi}}
\newcommand{\fmerg}{f_{1}}
\newcommand{\fring}{f_{2}}
\newcommand{\fcut}{f_{3}}
\newcommand{\rmi}{{\rm i}}
\def\rd{{\textrm{\mbox{\tiny{RD}}}}}
\def\qnr{{\textrm{\mbox{\tiny{QNR}}}}}
\newcommand{\A}{\mathcal{A}}
\newcommand{\NS}{\mathrm{NS}}
\newcommand{\CC}{\mathcal{C}}
\newcommand{\hr}{h_{\rm resp}}
\newcommand{\fhr}{\tilde h_{\rm resp}}
\newcommand{\inner}[2]{{\left\langle #1 \middle| #2 \right\rangle}}

\newcommand{\mdh}[1]{\textcolor{red}{\textit{Mark: #1}}}
\newcommand{\patricia}[1]{\textcolor{blue}{\textit{Patricia: #1}}}
\newcommand{\frank}[1]{\textcolor[rgb]{0,0.4,0}{\textit{(Frank: #1)}}}

\renewcommand{\Re}{\operatorname{Re}}
\renewcommand{\Im}{\operatorname{Im}}

\newcommand{\Cardiff}{School of Physics and Astronomy, Cardiff University, Queens Building, CF24 3AA, Cardiff, United Kingdom}


\title{Towards models of gravitational waveforms from generic binaries II: \\ 
Modelling precession effects with a single effective precession parameter}

\author{Patricia Schmidt}
\affiliation{\Cardiff}

\author{Frank Ohme}
\affiliation{\Cardiff}

\author{Mark Hannam}
\affiliation{\Cardiff}

\begin{abstract}

Gravitational waves (GWs) emitted by generic black-hole binaries show a
rich structure that directly reflects the complex dynamics introduced by the precession of
the orbital plane, which poses a real challenge to the development of generic waveform
models. Recent progress in modelling these signals relies on an
approximate decoupling between the non-precessing secular inspiral and a
precession-induced rotation. However, the latter depends in general on all
physical parameters of the binary which makes modelling efforts as well as
understanding parameter-estimation prospects prohibitively complex. Here we show
that the dominant precession effects can be captured by a reduced set of
spin parameters. Specifically, we introduce a single \emph{effective precession spin}
parameter, $\chi_p$, which is defined from the spin components that lie in the
orbital plane at some (arbitrary) instant during the inspiral. We test the
efficacy of this parameter by considering binary inspiral configurations specified by the
physical parameters of a corresponding non-precessing-binary configuration
(total mass, mass ratio, and spin components (anti-)parallel to the orbital
angular momentum), plus the effective precession spin applied to the larger
black hole. We show that for an overwhelming majority of random precessing
configurations, the precession dynamics during the inspiral are well
approximated by our equivalent configurations. Our results suggest that in the
comparable-mass regime waveform models with only three spin parameters
faithfully represent generic waveforms, which has practical implications for the
prospects of GW searches, parameter estimation and the numerical exploration of
the precessing-binary parameter space.
\end{abstract}


\maketitle

\begin{acronym}
\acrodef{BH}[BH]{black hole}
\acrodef{GW}[GW]{gravitational-wave}
\acrodef{NS}[NS]{neutron star}
\acrodef{PN}[PN]{post-Newtonian}
\acrodef{SNR}[SNR]{signal-to-noise ratio}
\acrodef{NR}[NR]{numerical relativity}
\acrodef{EOB}[EOB]{effective-one-body}
\acrodef{IMR}[IMR]{inspiral-merger-ringdown}
\acrodef{CDF}[CDF]{cumulative distribution function}
\end{acronym}

\newcommand{\GW}{\ac{GW}\xspace}
\newcommand{\PN}{\ac{PN}\xspace}
\newcommand{\NR}{\ac{NR}\xspace}
\newcommand{\EOB}{\ac{EOB}\xspace}
\newcommand{\IMR}{\ac{IMR}\xspace}
\newcommand{\CDF}{\ac{CDF}\xspace}

\section{Introduction}
\label{sec:intro}
The commissioning of the advanced interferometric \GW 
detectors is currently underway, with Advanced LIGO (aLIGO) \cite{Waldman:2011vg,
Harry:2010zz} expected to go online in late 2015, and to reach its
anticipated design sensitivity by 2019~\cite{Aasi:2013wya}. Advanced 
Virgo~\cite{AdVirgoDesign} in Italy and Kagra~\cite{Somiya:2011me} in Japan are
expected to follow soon after. The coalescence of two compact objects like black
holes or neutron stars is among the most promising candidates for the first
direct detection of \acp{GW}. However, the prime detection strategy for \acp{GW}
from coalescing compact binaries exploited by the ground-based detectors,
matched filtering, relies on theoretical knowledge of the gravitational
waveforms. It is therefore crucial to have accurate and efficient waveform models of the \GW
signal from binary coalescences readily available to use in the
advanced-detector era. 

The dynamical evolution of a compact binary system can be separated into three
distinct stages: the inspiral, the merger and the ringdown, if the final object is a 
black hole. Whilst inspiral
waveforms can be predicted accurately by analytic approximation methods like \PN
expansions (see \cite{lrr-2014-2} and references therein) or their resummation
into \EOB models \cite{Damour:1997ub,Buonanno:1998gg,Buonanno:2000ef}, the later
stages need to be calculated from \NR solutions of the full nonlinear Einstein field equations.
Current waveform models for the complete \IMR \GW signal were constructed by combining
analytic \PN/\EOB and \NR results. 

To date, a number of such theoretical IMR waveform models exist for nonspinning
black-hole
binaries~\cite{Ajith:2007qp,Ajith:2007kx,Buonanno:2007pf,Buonanno:2009qa,
Damour:2007vq,Damour:2008te,Damour:2009kr,Pan:2011gk,Damour:2012ky} as well as
for spinning binary configurations where the orbital angular momentum of the
binary motion is (anti)-parallel to the spin angular momenta of the individual
holes~\cite{Ajith:2009bn,Pan:2009wj,Santamaria:2010yb,Taracchini:2012ig} (see
also \cite{Ohme:2011rm} for an overview). In these configurations the black
holes orbit in a spatially fixed two-dimensional plane, and the dominant mode of the \GW signal can be
described by simple monotonic functions for the amplitude and phase. The spins modify the
inspiral rate and the \GW amplitude, but otherwise the \GW signals are
qualitatively similar between non-spinning and aligned-spin configurations. 
The \NR simulations needed to calibrate these \IMR models cover a three-dimensional
parameter space of the binary's mass ratio and the two spin magnitudes (the
total mass is a simple scaling factor for vacuum solutions). The models produced
to date were calibrated with $\sim30$ \NR simulations, c.f.\
Refs.~\cite{Ajith:2009bn,Santamaria:2010yb}
and Ref.~\cite{Taracchini:2013rva}.

In the most general cases, however, the individual spin angular momenta have arbitrary
orientations, 
and any misalignment between the orbital angular momentum and the spins causes
the orbital plane as well as the spin vectors to
precess~\cite{Apostolatos:1994mx,Kidder:1995zr}. The simple inspiral motion
becomes more complicated, leading to a \GW signal with 
amplitude and phase modulations that depend on the orbital-plane orientation, 
as well as a richer mode structure. We will discuss in
more detail the phenomenology of precessing binaries in Sec.~\ref{sec:phen}.

During the last four years, a number of key results have helped to develop a
simple framework to model the waveforms of precessing black-hole binaries
\cite{Schmidt:2010it, Boyle:2011gg, O'Shaughnessy:2012vm, Schmidt:2012rh,  
Pekowsky:2013ska}.
In particular, in earlier work we showed that the waveform from the inspiral of a
precessing binary can
be approximated by an underlying non-precessing-binary waveform that has been
``twisted up'' by the precessional motion of the orbital plane
\cite{Schmidt:2012rh} (see Fig. 5 of Ref.~\cite{Schmidt:2012rh}). The non-precessing waveform is characterised by the
individual masses and the components of the black-hole spins parallel
to the orbital angular momentum, which remain roughly constant throughout the
inspiral.
Our results indicated that this mapping holds up to the merger; later work quantified that the merger and ringdown can also be mapped
to non-precessing-binary waveforms, but the parameter identification is less
clear~\cite{Pekowsky:2013ska}. 
The inspiral dynamics (predominantly influenced by the
individual masses and the ``parallel'' spin components) can be approximately decoupled from the
precession, which is determined by \emph{all} of the physical parameters, and
this suggests an elegant way to construct a generic-binary model, i.e., to
separately produce a waveform model for aligned-spin binaries and an additional 
model for the precessional motion~\cite{Schmidt:2012rh}. This proposal has since
been exploited to construct precessing \IMR models~\cite{Pan:2013rra,Hannam:2013oca}.

An open problem in modelling generic binaries (i.e., including precession
effects in the merger and ringdown) is the need for \NR simulations across
a seven-dimensional parameter space (mass ratio, plus the vector components of
each black hole's spin). \NR simulations are computationally expensive, and even
a coarse sampling of four points in each direction of the parameter space would
require $4^7 \sim O(10^5)$ simulations. One way to make this problem tractable
is to identify the physical parameters (or combinations of them) that most
strongly affect the \GW signal. This approach will not only provide us with a
smaller subspace over which to perform \NR simulations, but will also indicate those
physical parameters that can most accurately be measured in future \GW observations. 

This approach has already been used in some models of spinning, non-precessing binaries:
the spins predominantly affect the inspiral rate, but this influence can be parameterized by
a weighted sum of the two spins, and therefore efficient aligned-spin models can
be produced with only \emph{one} spin parameter rather than
two~\cite{Ajith:2009bn,Santamaria:2010yb,Ajith:2011ec,Reisswig:2009vc,Purrer:2013ojf}.
Our goal in this work is to identify a
complementary spin parameter for precession and reduce the remaining four dimensions
(the in-plane spin components) to a subspace that accurately captures the dominant
precession-induced features in \GW signals across the full parameter space. We
find that a single additional ``precession spin parameter", which we denote $\chi_p$, is
sufficient for this purpose, and we investigate its efficacy in a study of \PN
inspiral waveforms for generic comparable-mass-ratio binaries. 

Preliminary work on this effective precession spin parameter motivated the choice of
parameters in our phenomenological frequency-domain \IMR model,
PhenomP~\cite{Hannam:2013oca}. This work also provides additional justification
for single-spin waveform models, such as the Physical Template Family
\cite{Buonanno:2004yd} and the precessing stationary-phase inspiral model
 in Ref.~\cite{Lundgren:2013jla}.

This paper is organised as follows. In Sec.~\ref{sec:phen} we
briefly summarise the phenomenology of simply precessing binaries and recent developments 
in modelling precessing binaries. In Sec.~\ref{sec:chip} we introduce the effective precession spin parameter
$\chi_p$. In Sec.~\ref{sec:pn} we describe the \PN waveforms and analysis procedure we use to 
quantify the accuracy of waveforms where the in-plane spins are mapped to $\chi_p$, and the results
are presented in Sec.~\ref{sec:results}. Based on these we discuss the applicability of the 
$\chi_p$ approximation in Sec.~\ref{sec:conclusion}.

\section{Precessing black-hole binaries}
\label{sec:phen}

\subsection{Phenomenology}
We briefly summarise the essential features of precession and its effects on the
\GW signal. For a more detailed
discussion we refer the reader to Refs.~\cite{Apostolatos:1994mx,Kidder:1995zr}.

The loss of binding energy via \acp{GW} causes two Kerr black holes with
component masses $m_1$ and $m_2$ in a quasi-circular orbit to spiral inwards 
until they merge into a single black hole. If the black holes' spin angular
momenta $\vec S_i$ are aligned (anti-)parallel to the 
orbital angular momentum $\vec{L}$, then the orbital motion occurs in a fixed two-dimensional plane, defined by
$\hat{L}$, which is also the direction of dominant GW energy emission. 

This simple picture changes when the individual spins have some arbitrary orientation. In such generic 
configurations, the orientations of the individual spins and the orbital plane evolve. In most configurations 
the binary follows \emph{simple precession}, where both the spin and orbital angular momenta precess 
around the binary's total angular momentum, $\vec{J}= \vec{L}+\vec{S}_1+\vec{S}_2$~\cite{Apostolatos:1994mx}. 
The direction of the total angular momentum is approximately fixed, i.e., $\hat{J}(t) \simeq \hat{J}_{t\rightarrow -\infty}$, 
and is therefore a natural generalisation of the orbital angular momentum as characteristic direction 
in the binary system. If $\hat{N}$ is the line-of-sight direction of a distant inertial observer (detector), then we can define 
$\theta=\measuredangle (\hat{J},\hat{N})$ as the inclination of the binary.

When $L \simeq S$ and $\hat{L} \sim -\hat{S}$, then small changes in $J$ due to
GW emission are comparable to 
the magnitude of $J$, and its direction is {\it not} fixed; on the contrary it ``tumbles over'' (see Fig. 6 in Ref.~\cite{Schmidt:2012rh}). This is called \emph{transitional precession}.
Only a very restricted set of physical configurations will undergo transitional precession while emitting GWs at
frequencies within the sensitivity band of the Advanced GW detectors, and therefore observations of these systems
are expected to be rare~\cite{Apostolatos:1994mx}. 

\begin{figure}
\def\svgwidth{0.8\columnwidth}
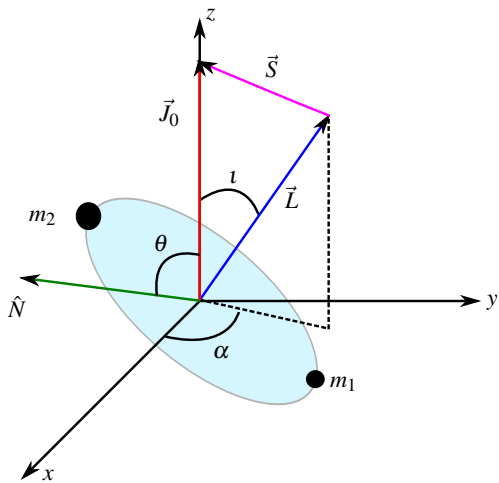
\caption{The $\hat{J}_0$-aligned source frame of a precessing binary. $\theta$ denotes
the angle between the line of sight (radiation propagation direction) and the
total angular momentum; $\vec{S}=\vec{S}_1+\vec{S}_2$ is the total spin.}
\label{fig:SourceFrame}
\end{figure}

In the following, we adopt a Cartesian coordinate system attached to the binary
such that at the initial time $\hat{J}_0 \equiv \hat{z}$, which we refer to as the $J_0$-aligned
source frame. Therein, we define the instantaneous direction of the orbital angular
momentum, $\hat{L}(t)$, by the two polar angles $(\iota(t), \alpha(t))$. These
functions encode the time evolution of the orientation of the orbital plane in
the source frame. The precession cone opening angle $\iota(t)$ is defined by
\begin{equation}
\label{eq:iota}
\iota(t):= \arccos{\left(\hat{L}(t)\cdot \hat{J}(t)\right)},
\end{equation}
and the azimuthal angle $\alpha(t)$ is given by
\begin{equation}
\label{eq:alpha}
\alpha(t):=\arctan \left(\frac{L_y}{L_x}\right).
\end{equation}
The geometry of a precessing configuration is depicted in Fig.~\ref{fig:SourceFrame}. Due to its nature, the azimuth angle is directly related to the precession frequency, i.e., the rate at which $\hat{L}$ precesses around $\hat{J}$,
\begin{equation}
\label{eq:omegap}
\omega_p(t)=\frac{d\alpha(t)}{dt}.
\end{equation}

\begin{figure}
\begin{center}
\includegraphics[width=80mm]{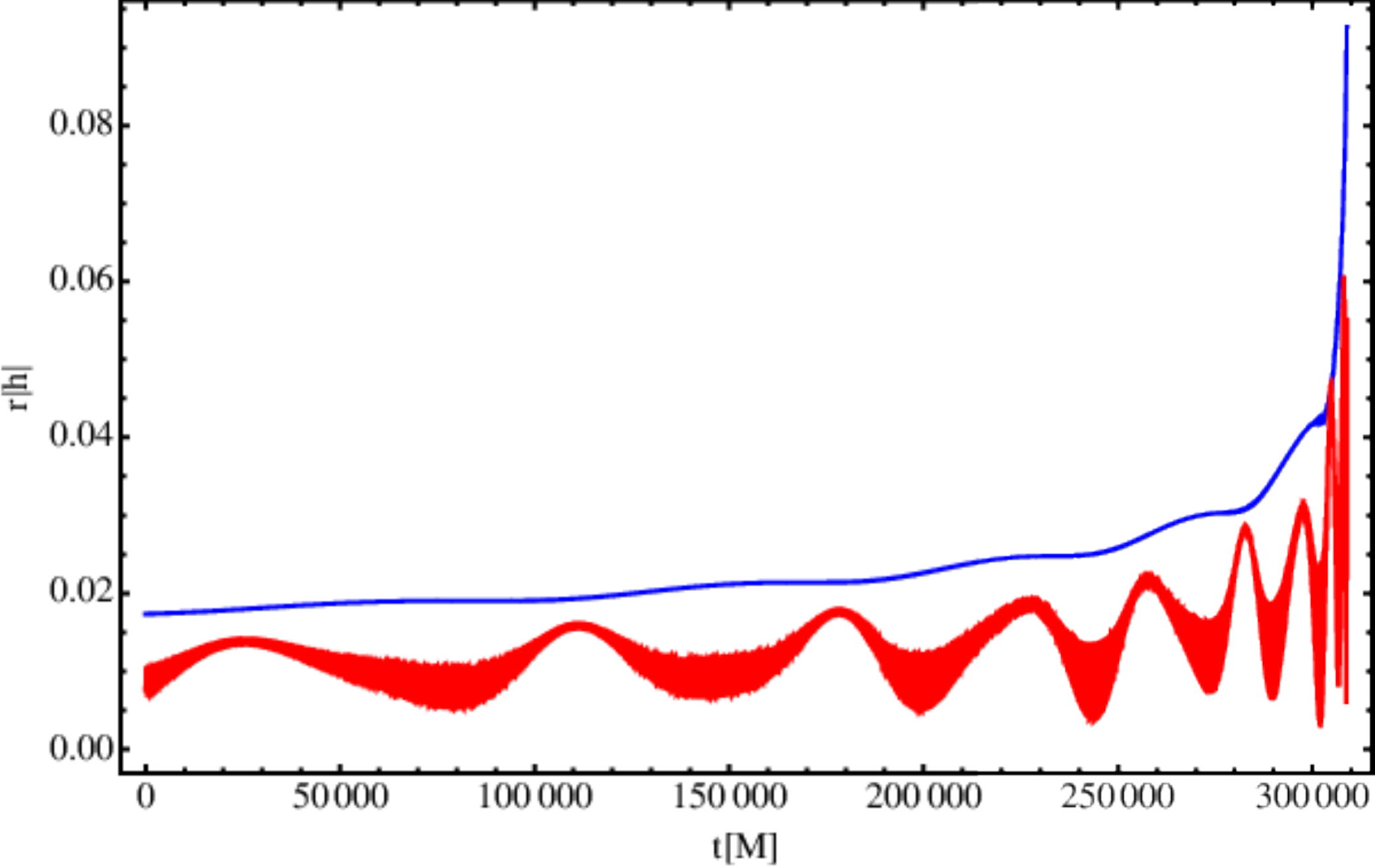}
\caption{Magnitude of the GW strain $h$ computed with all $\ell=2$ modes for a
precessing binary, where the total angular momentum $\hat{J}_0$ is aligned with
the line-of-sight (top blue curve) and for the arbitrary orientation $(\theta,
\phi)=(60^\circ, 113^\circ)$ (lower red curve). The binary's parameters are
$q=3$, $\vec{\chi}_1=(1,0,0)$ and $\vec{\chi}_2=(0.8,0,0.6)$. While only weak
amplitude modulations are visible along $\hat{J}_0$, we observe strong
modulations for the arbitrary orientations.
}
\label{fig:hJ0}
\end{center}
\end{figure}

Precession occurs due to spin-orbit and spin-spin couplings and is a purely relativistic effect. In PN theory,
the evolution of $L$ and $S_i$ can be described through 2.5PN order by the precession equations 
Eq.~(\ref{eq:Ldot})-Eq.~(\ref{eq:S2dot}). The leading-order effect occurs at 1.5-PN order (spin-orbit coupling) 
and drives the precession of the orbital plane (Lense-Thirring precession); the dominant spin-spin coupling 
term appears at 2PN order and induces nutational motion. The precession of the
orbital plane and the spins alter the otherwise simple orbital motion and
consequently affect the \GW emission. Most importantly, precession introduces 
a secular modification to the signal phase $\Phi(t)$, given by 
\begin{equation}
\label{eq:totalphase}
\Phi(t)=\int_0^t (\omega_\mathrm{orb}(t')-\dot{\alpha}(t')\cos \iota(t'))dt',
\end{equation}
as well as amplitude and phase modulations, and in the relative amplitudes of the waveform modes.
We emphasise that the strength of the modulations depends strongly on the relative orientation 
of the binary to the observer, i.e., $\theta$. Even strongly-precessing systems can show only mild modulations 
if the the observer is aligned with $\hat{J}_0$, i.e., $\theta=0$. The effect of the orientation on the  modulations 
is illustrated in Fig.~\ref{fig:hJ0}. 

\subsection{Modelling simple precession: summary and recent progress}
\label{sec:summary}
Since accurate waveform templates are a key
ingredient in most detection and parameter-estimation strategies, finding
accurate and efficient waveform models for generic binaries has been an
ongoing challenge for several decades. In the inspiral regime this has meant
finding simple ways to capture the dominant precession effects, without having
to solve the full \PN or \EOB equations of motion, which is prohibitively expensive
in GW applications. Here we give a
brief overview of the most important recent developments but refer to
Ref.~\cite{Hannam:2013pra} for a more complete treatment. 

First attempts to construct search templates for precessing
signals~\cite{Apostolatos:1995pj, Apostolatos:1996rf} followed soon after the
analysis of the phenomenology of precessing binaries within the \PN framework by
Apostolatos et al.~\cite{Apostolatos:1994mx} and Kidder~\cite{Kidder:1995zr}.
Apostolatos was the first to observe the potential of modulating the
\emph{secular phase}, which he referred to as the ``carrier phase'', to describe
the total phase of the precessing system. Schematically, the precessing GW 
strain $h$ is then given as
\begin{equation}
h(t) = \Lambda(t)h_C(t),
\end{equation}
where $h_C(t)$ is the unmodulated carrier signal and $\Lambda(t)$ is a complex
factor containing all information regarding the precession-induced modulations
of the amplitude and the phase (see Eq.(6)-Eq.(17) in Ref.~\cite{Apostolatos:1995pj}
for details). Crucially, this ansatz assumes that the unmodulated carrier phase
is that of a \emph{nonspinning binary}. 
Apostolatos concluded that the agreement between the artificially modulated waveforms 
and true precessing waveforms is unacceptably low even for moderate
precession~\cite{Apostolatos:1995pj}. 

Subsequently, Buonanno, Chen and Vallisneri~\cite{Buonanno:2002fy} (BCV)
modified the modulation factor in Apostolatos' general ansatz. However, the
description of the secular phase was unchanged. The modified $\Lambda(t)$ was
able to capture the precession-induced modulations better, but in order to do
so, up to six free non-physical parameters were introduced, which were
subsequently shown to admit waveforms that mimicked detector noise and lead to
an increase in the false alarm rate of a GW search~\cite{VanDenBroeck:2009gd}. 

In previous work~\cite{Schmidt:2012rh}, we have suggested to model \acp{GW} from
generic black-hole binaries in a similar way, but we identified the carrier
signal with an appropriate aligned-spin waveform which is ``twisted up''
following the precession dynamics. We proposed
\begin{align}
    h^\mathrm{nonspinning}_C(t)& \rightarrow h^\mathrm{nonprecessing}(t),    \\
    \Lambda(t)& \rightarrow \mathbf{R}(t), 
\end{align}
where the modulation factor $\Lambda$ becomes a simple rotation operator
$\mathbf{R}$ with a concrete physical meaning: it encodes the evolution of the
orbital plane. 

Whilst aligned-spin binaries have been accurately modelled in the past, the
missing ingredient is a sufficient description of the rotation operator $\mathbf{R}$, which depends
on the precession angles $\iota$ and $\alpha$. Exact solutions to the leading-order PN precession 
equations are known for 
two special cases, equal-mass or single-spin binaries~\cite{Apostolatos:1994mx},
but in general analytic solutions are not known in the comparable mass
regime.
In addition, the angles $\alpha$ and $\iota$ depend in general on all six spin components, which
significantly complicates modelling efforts. In order to establish a sufficiently
accurate but simple model for the two angle functions,
it would be advantageous to reduce the number of
dependent parameters, and we shall motivate a single parameter that
governs the precession dynamics in the following section.

\section{Effective precession spin}
\label{sec:chip}

Generic binary black holes are in general characterised by seven intrinsic
physical parameters: the mass ratio $q=m_2/m_1\geq 1$, and the six spin
components of their two spin angular momenta 
$\vec{S}_i$ ($i=1,2$), or their dimensionless counterparts $\vec
\chi_i=\vec S_i/m_i^2$. The total mass of the binary 
sets the overall scale in General Relativity and therefore need not be
explicitly included in a waveform model. 

\begin{figure}
\includegraphics[width=80mm]{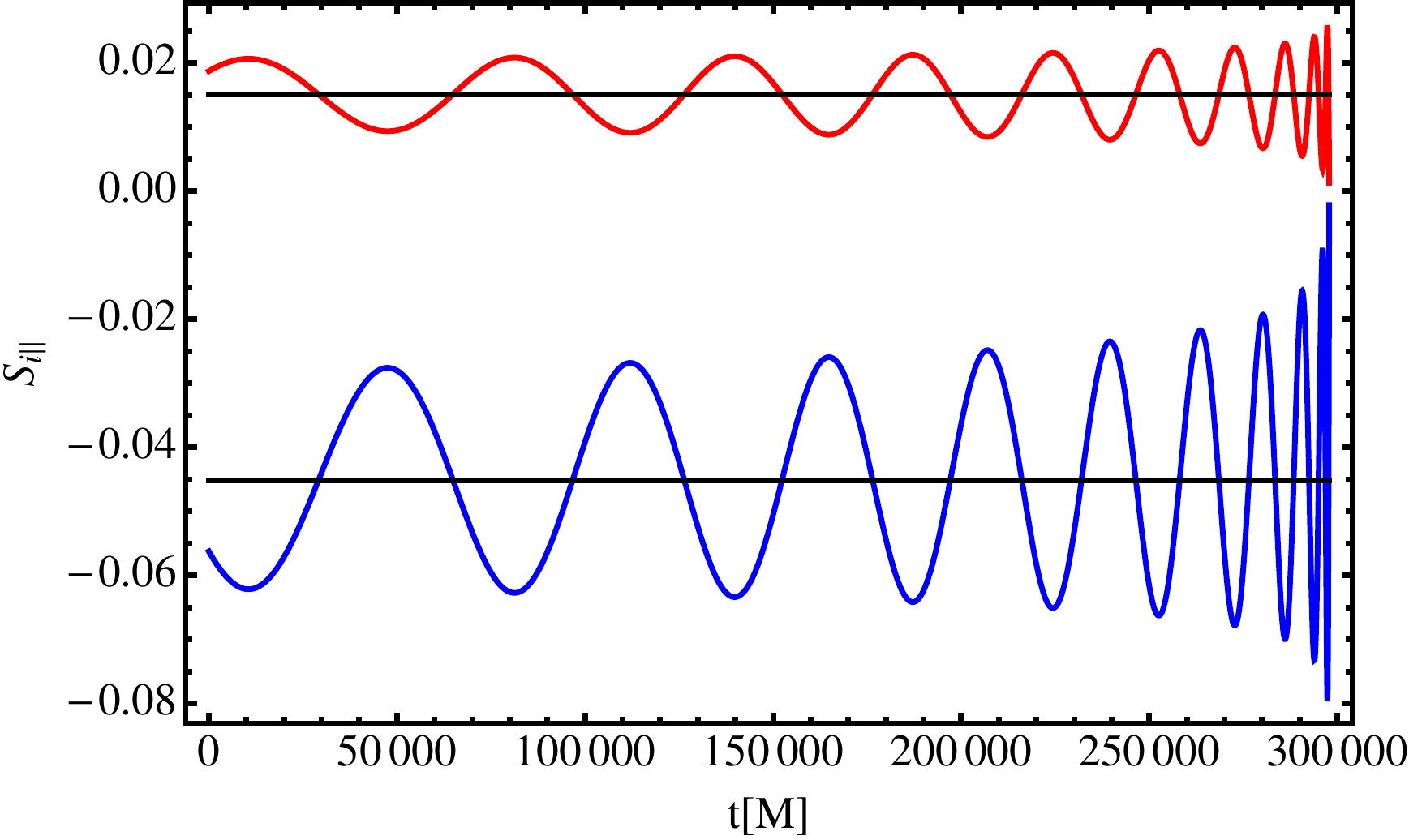}
\caption{Evolution of the two spins parallel to the orbital
angular momentum. The top red graph shows the evolution of the parallel spin of
the smaller black hole, $S_{1||}$, the lower blue curve that of the parallel
spin of the larger black hole, $S_{2||}$ for the case described in the text. The
two horizontal lines indicate the mean value of each parallel spin with
$\bar{S}_{1||}=0.015$ and $\bar{S}_{2||}=-0.045$.}
\label{fig:Spara}
\end{figure}

\begin{figure}
\begin{center}
\includegraphics[width=75mm]{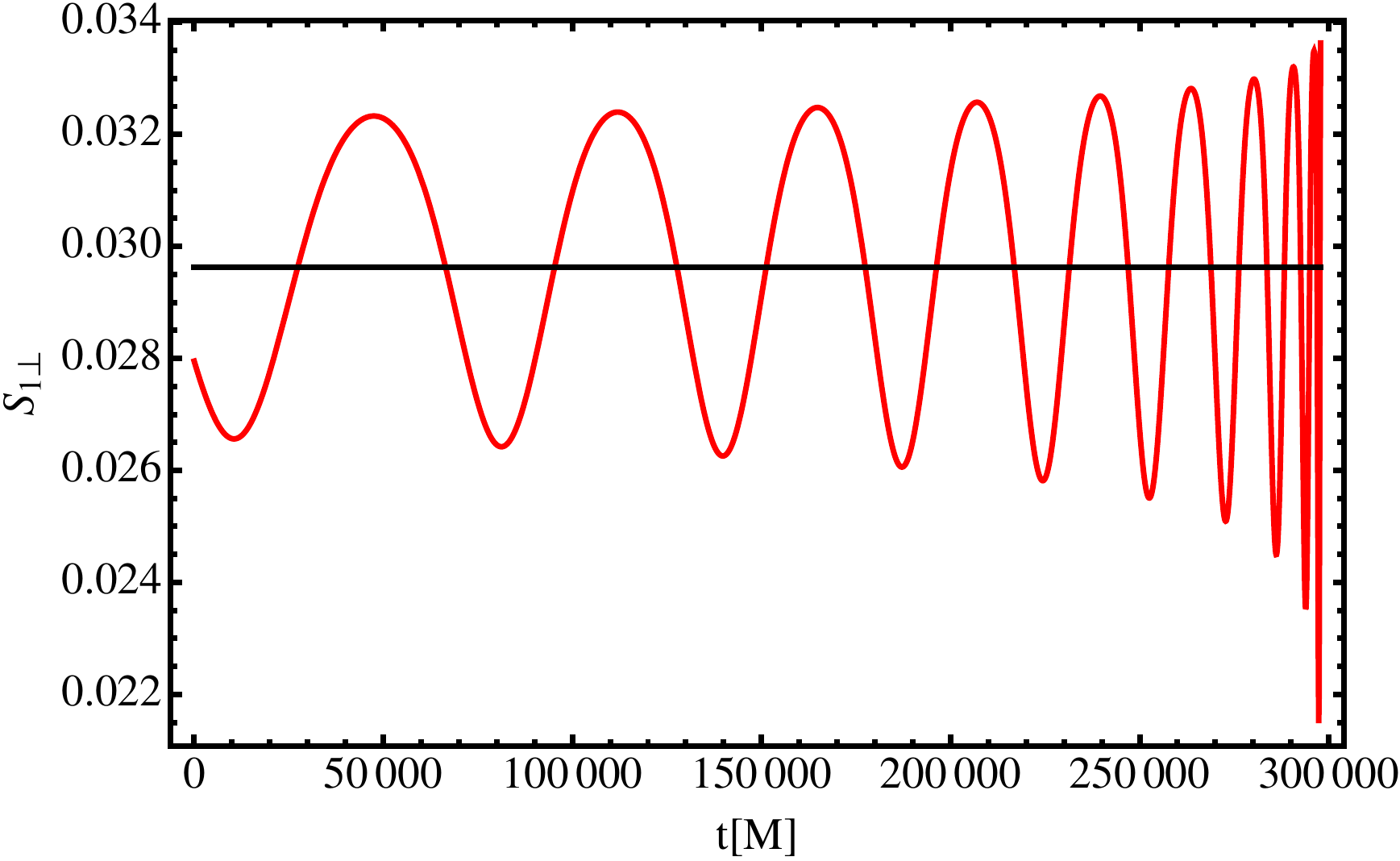}
\includegraphics[width=75mm]{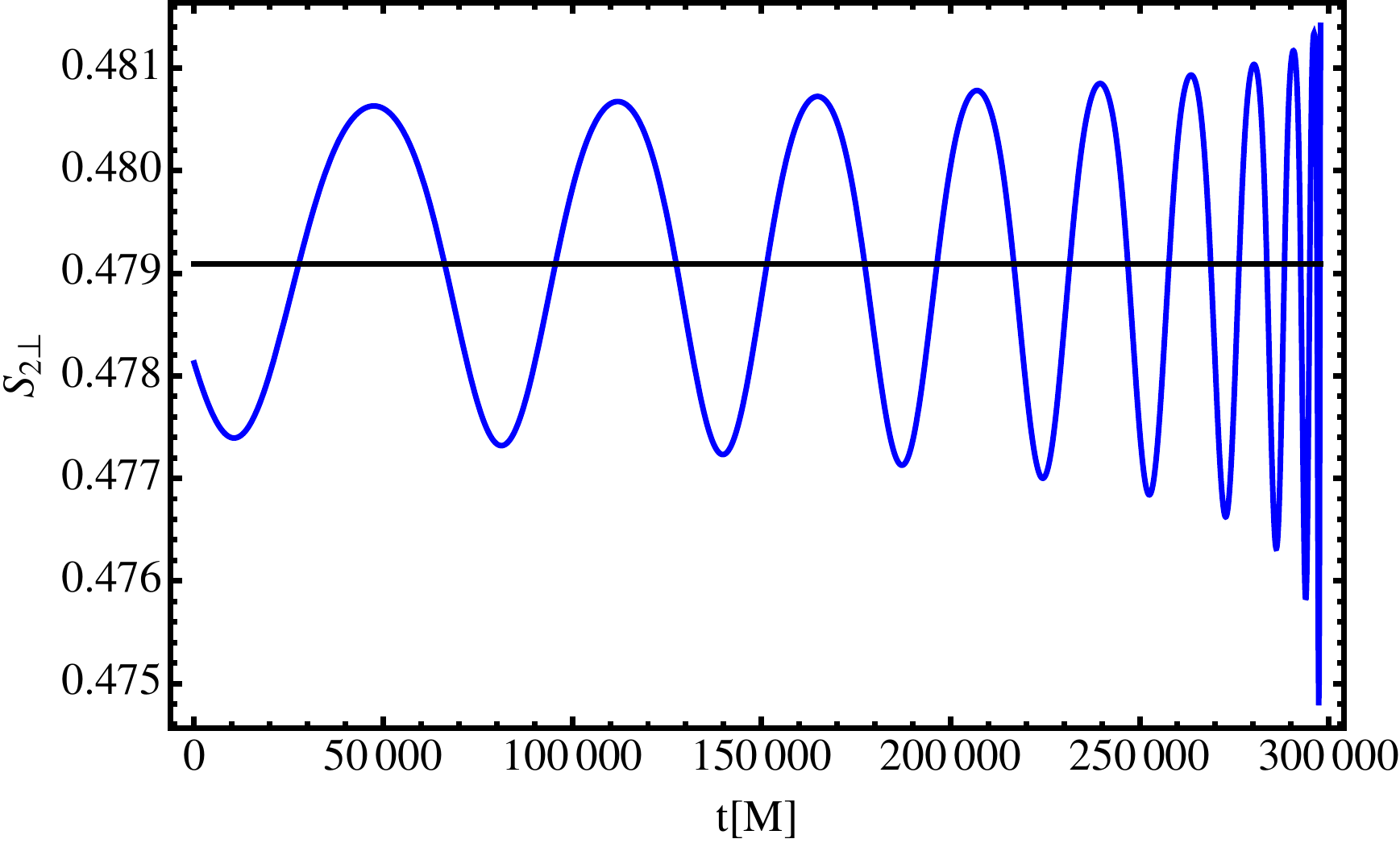}
\caption{The top panel shows the evolution of
$S_{1\perp}$ as function of time, the bottom panel shows
the evolution of $S_{2\perp}$. Similar to the parallel
spin magnitudes, the in-plane spin magnitudes oscillate around some mean
values, which are $\bar{S}_{1\perp}=0.030$ and $\bar{S}_{2\perp}=0.479$
respectively (horizontal lines).}
\label{fig:Sortho}
\end{center}
\end{figure}

In previous work \cite{Schmidt:2012rh, Pekowsky:2013ska}, it was shown that the
secular phasing, i.e., the inspiral rate, of precessing binaries is determined
by the mass ratio and spin components parallel to the orbital angular momentum,
$S_{i\parallel}=\vec{S}_{i}\cdot \hat{L}$. These are approximately constant, in 
that they exhibit only small variations 
throughout the inspiral, even for generic binaries. This behaviour is
illustrated in Fig.~\ref{fig:Spara} for a precessing binary with mass ratio
$q=3$, $\vec{\chi}_1=(0.4,-0.2,0.3)$ and $\vec{\chi}_2=(0.75,0.4,-0.1)$. 
(The details of our PN waveform generation are given in Appendix~\ref{sec:wfgen}.)
We see
that the parallel spin components each oscillate around a mean value, which is
close to the initial values of $S_{1||}=0.01875$ and $S_{2||}=-0.05625$. 
Note that
the individual total spin magnitudes $S_i$ are conserved, and the observed
oscillations in the parallel spin magnitudes are compensated by changes in the
in-plane spin magnitudes at each moment in time which in turn is illustrated in
Fig.~\ref{fig:Sortho}. We note that these oscillations occur on the precession
and not the orbital timescale, and, once again, the in-plane spin magnitudes
oscillate around a approximately fixed mean values. For comparison, in this 
case the initial in-plane magnitudes were 0.0279 and 0.478.

To describe the precession, we require additional information from the spin components that lie in the orbital plane orthogonal to $\hat{L}$.   It is therefore convenient to decompose the spin vectors with respect to $\hat{L}$ into their parallel and orthogonal vector components such that each spin vector $\vec{S}_i=\vec{S}_{i||}+\vec{S}_{i\perp}$. In the following, however, we will show that it is possible to faithfully
approximate the precession in a generic binary system by combining these four 
in-plane spin components $\vec{S}_{1\perp}$ and $\vec{S}_{2\perp}$ into only one additional spin parameter, a
complementary \emph{effective precession spin}, $\chi_p$.

Consider the leading-order \PN precession equation~\cite{Apostolatos:1994mx}:
\begin{align}
\dot{\vec{L}}  &=  \frac{L}{r^3}\left[ \left ( 2
+\frac{3q}{2}\right)\vec{S}_1+ \left( 2+\frac{3}{2q}\right) \vec{S}_2\right]
\times \hat{L} \\
& \equiv \frac{L}{r^3}\left[ A_1 \vec{S}_{1\perp} + A_2 \vec{S}_{2\perp} \right] 
\times \hat{L},
\label{eq:Ldot2}
\end{align}
where $A_1=2+3q/2$ and $A_2=2+3/(2q)$, and $r$ denotes the separation. We see immediately that the in-plane spins $\vec{S}_{i\perp}$ drive the evolution of $L$. Similar evolution equations are given 
for the spin vectors (see Eq.(\ref{eq:S1dot})-Eq.(\ref{eq:S2dot})). At leading order these suggest 
that the in-plane spins $\vec{S}_{i\perp}$ rotate within the orbital plane, but with different rotational velocities, i.e., they have
different precession rates around $\hat{L}$. Their magnitudes $S_{i\perp}$ may also oscillate,
as shown in Fig.~\ref{fig:Sortho}, indicating the nutation of the orbital plane. The magnitude 
of these oscillations is typically small, and need not be modelled accurately in order describe the waveform 
faithfully (as quantified in Sec.~\ref{sec:pn}). Instead, in the following we focus on modelling the
\emph{average} precession of the orbital plane.

\begin{figure}
\centering
\includegraphics[width=80mm]{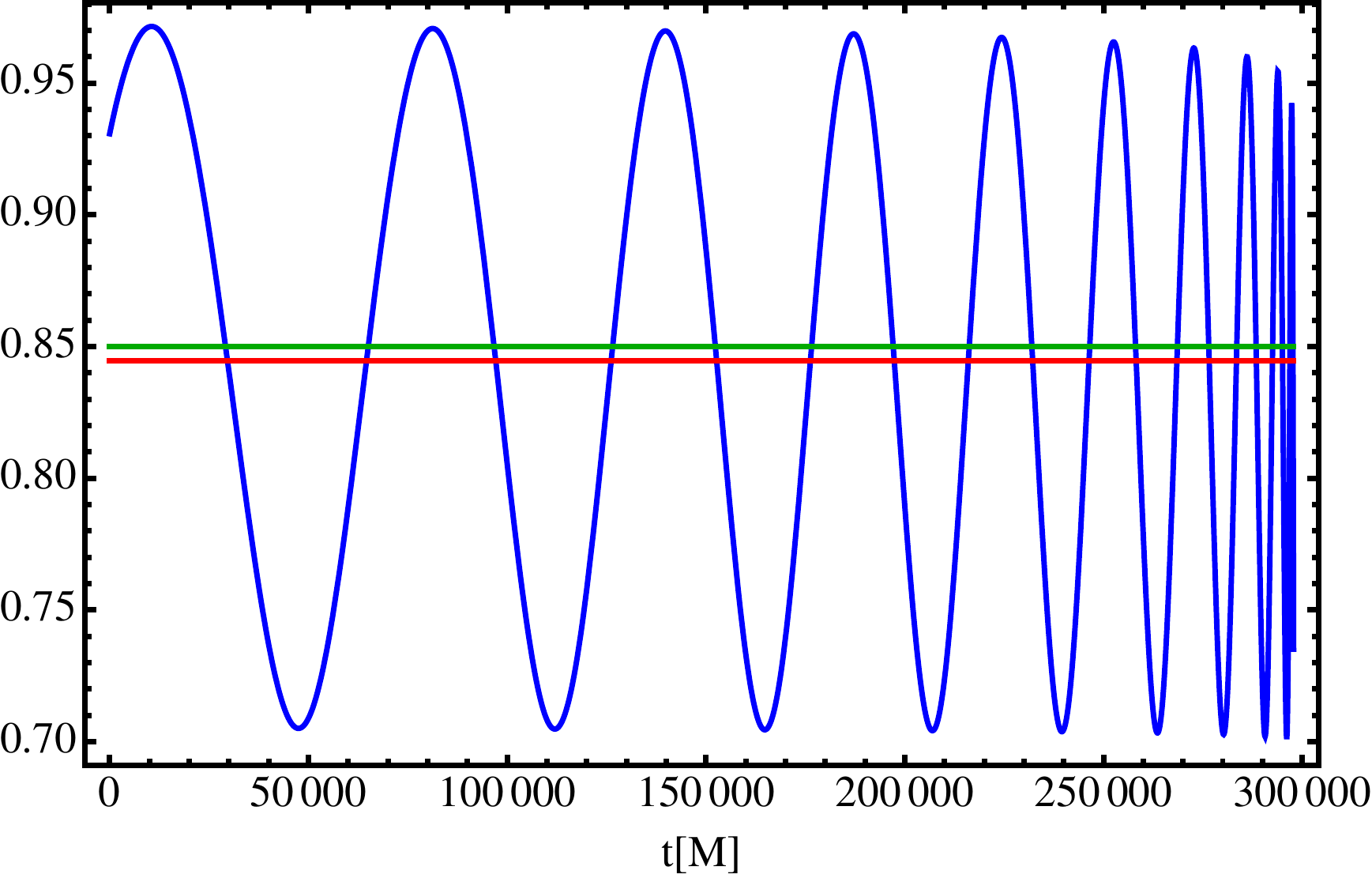}
\caption{Magnitude of the leading-order precession term $\| (A_1 \vec{S}_{1}
+ A_2 \vec{S}_{2})\times \hat{L} \|/(A_2 m_2^2)$ (blue), its true mean $\overline{\| (A_1
\vec{S}_{1} + A_2 \vec{S}_{2})\times \hat{L} \|/(A_2 m_2^2)}=0.845$ (lower red
horizontal line) and its approximation $\chi_p=0.85$ as determined from Eq.~(\ref{eq:chip}) 
(upper green horizontal
line).}
\label{fig:ChiP}
\end{figure}

The two observations we have just made, 1) that the magnitudes of
the in-plane spins $S_{i\perp}$ each oscillate around a mean value and 
2) that the relative angle between the spin vectors in the plane changes continuously,
suggest a simple way to construct a single precession spin parameter. At some
times during the inspiral, the two in-plane spin vectors will be parallel, and
will add together in Eq.~(\ref{eq:Ldot2}). At other times, the in-plane spin
vectors will point in opposite directions, and their contributions will be
minimised. Over many precession cycles, the overall contribution to
Eq.~(\ref{eq:Ldot2}) can be approximated by the average magnitude of these two
contributions:
\begin{align}
\label{eq:Sp}
    S_p &:= \frac{1}{2}\left(A_1S_{1\perp}+A_2S_{2\perp}+|A_1S_{1\perp}-A_2S_{2\perp}|\right)  \nonumber \\
    &\equiv \max(A_1 S_{1\perp},A_2 S_{2\perp}),
\end{align}
This parameter can be defined at any point during the inspiral, and the
variation from the true mean value will typically be small. This is illustrated
in Fig.~\ref{fig:ChiP}. We see that $S_p$ is directly related to the in-plane spin angular 
momentum of one of the black holes. As we will see below, in most configurations
this \emph{is} the in-plane spin of the larger black hole.  

We now use  $S_p$ to approximate all four in-plane spin parameters, and we
are free to distribute the precession spin appropriately between the two black
holes in the binary. Motivated by the fact that the in-plane spin of the smaller
black hole becomes more and more negligible with increasing mass ratio, we
assign the precession spin completely to the larger black hole, and define the
dimensionless precession spin parameter as
\begin{equation}
\label{eq:chip}
\chi_p := \frac{S_p}{A_2 m_2^2}.
\end{equation}
For a small subset of configurations $\chi_p$ does not respect the Kerr limit
of $\chi_i \leq 1$, i.e., when $S_{2||}$ and $S_{1\perp}$ are both large. However, we
find for the random sample of configurations studied in Sec.~\ref{sec:results} that this is 
rare: we find $\sim$3\% of such configurations for $q=1$ binaries, and none in our sample 
for $q=3$ and $q=10$. 

\begin{figure}
\begin{center}
\includegraphics[width=0.49\textwidth]{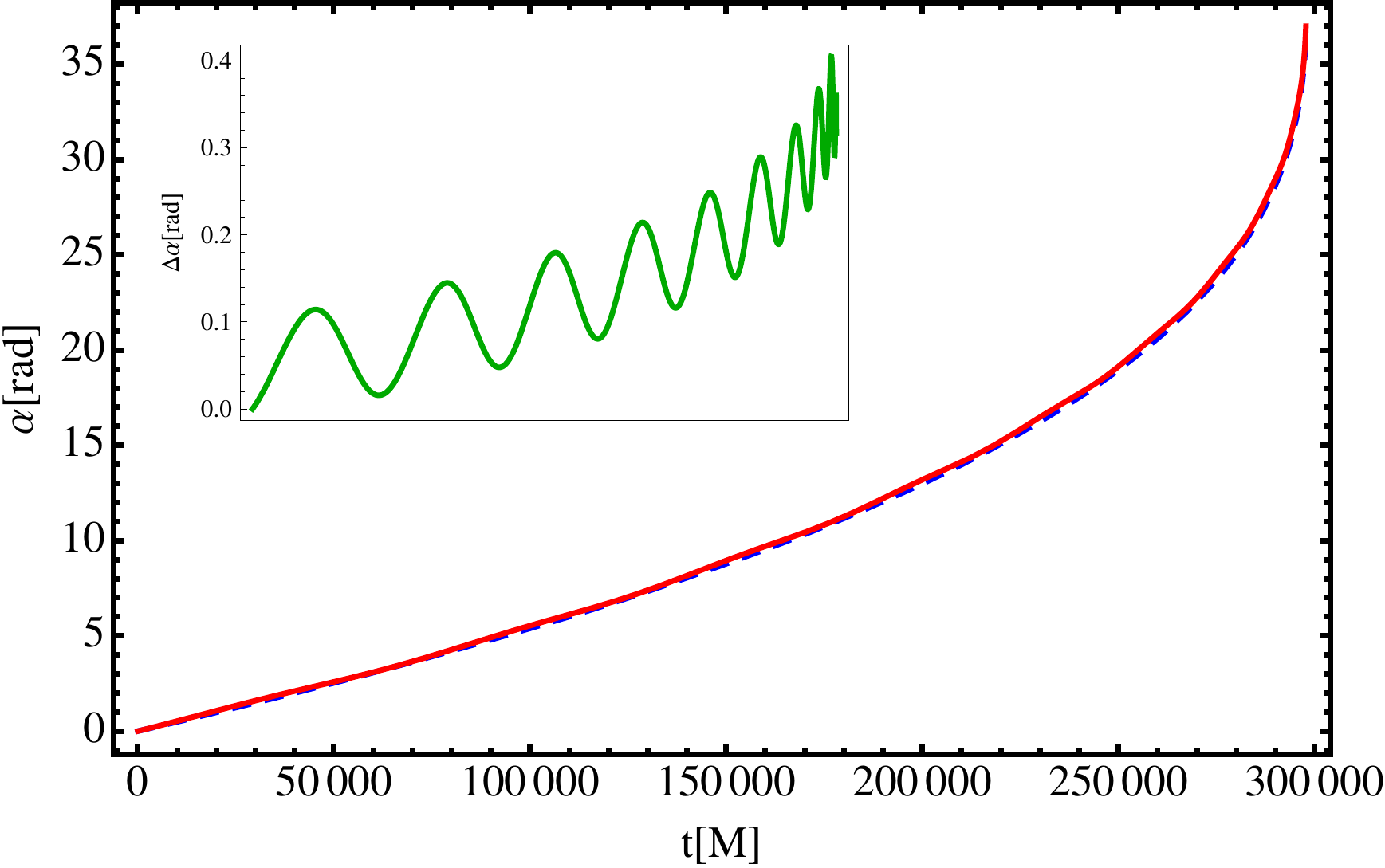}
\includegraphics[width=0.49\textwidth]{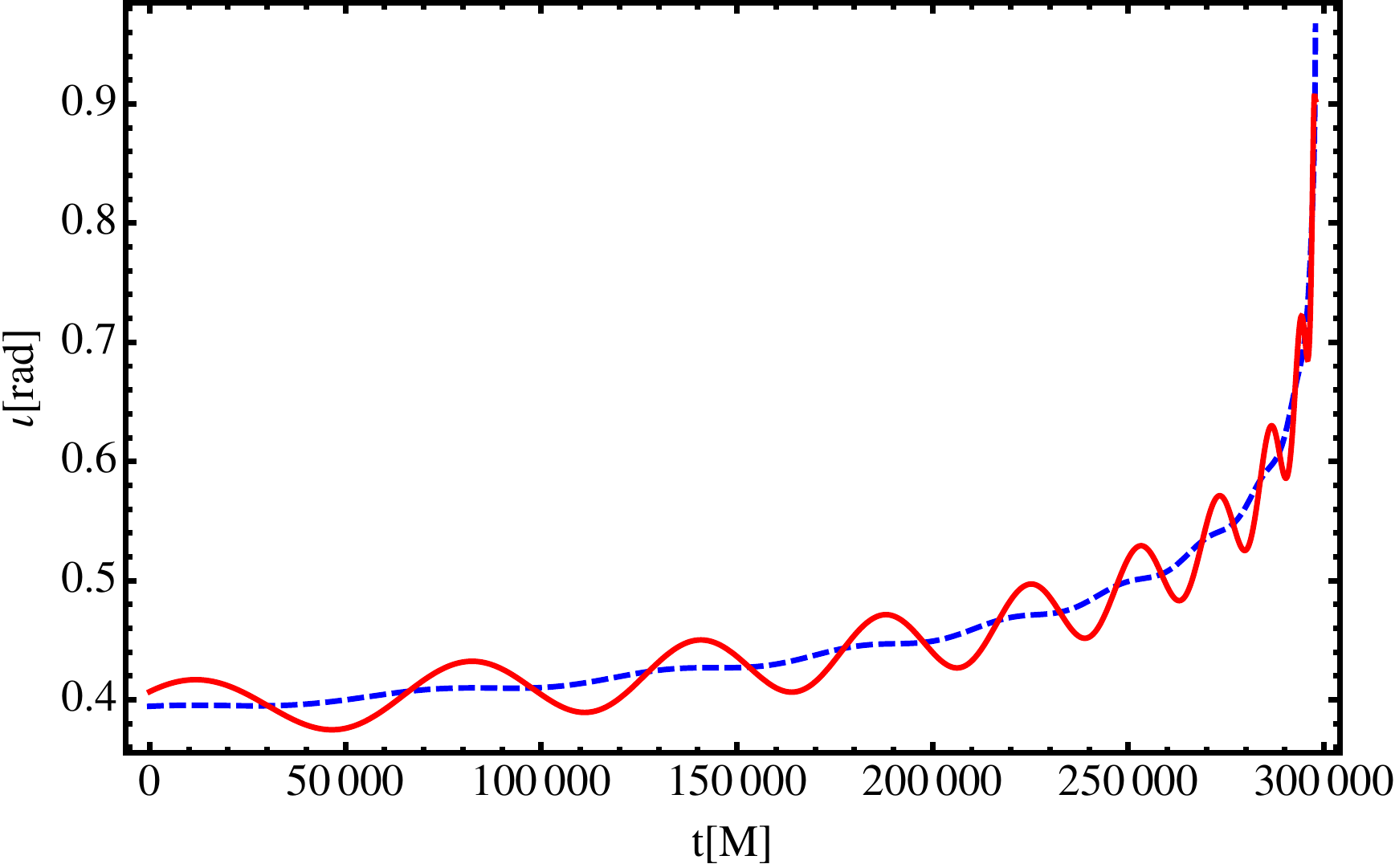}
\caption{The top panel shows $\alpha(t)$
for the generic configuration
$\{q=3,\vec{\chi}_1=(0.4,-0.2,0.3),\vec{\chi}_2=(0.75,0.4,-0.1)\}$ (red) and the
corresponding configuration utilising $\chi_p$ given by
$\{q=3,\vec{\chi}_1=(0.,0.,0.3),\vec{\chi}_2=(0.85,0.,-0.1)\}$ (blue, dashed)
Since the two curves are not distinguishable over that time scale, the inset
shows the difference $\Delta\alpha$ as a function of time. The bottom
panel compares the evolution of the opening angle of the precession cone
$\iota(t)$. Both graphs reveal that the approximation discards the spin-spin
couplings in the plane and therefore nutation effects (the visible
oscillations).}
\label{fig:AngleComp}
\end{center}
\end{figure}

Having chosen $\chi_p$ to be the approximate mean of the leading-order term in
the \PN precession equation, we expect (by construction) to see a similar
evolution of the orbital plane in a system where $\chi_p$ is used instead of
$S_{1\perp}$ and $S_{2\perp}$. Note that our definition of $\chi_p$ does \emph{not}
reproduce the same initial value of the precession cone opening angle, $\iota$; 
that would require that we instead focus on the average of $S_{1
\perp} + S_{2 \perp}$, and not the weighted sum in Eq.~(\ref{eq:Sp}). However,
we find the effect on $\iota$ to be small, and we also expect that it is less
important to correctly model $\iota$ than the precession angle $\alpha$ due it
its effect on the phase. This is illustrated for one generic case in
Fig.~\ref{fig:AngleComp}. We see that precession angles obtained from a
configuration, where the in-plane spins are replaced by $\chi_p$ on the larger
black hole, indeed represent the average precession of the full generic system.

There are two situations where we expect that applying a spin of $\chi_p$ to the
larger black hole may not adequately capture the average precessional motion of the
corresponding full system. 

\begin{figure}
\begin{center}
\includegraphics[width=0.45\textwidth]{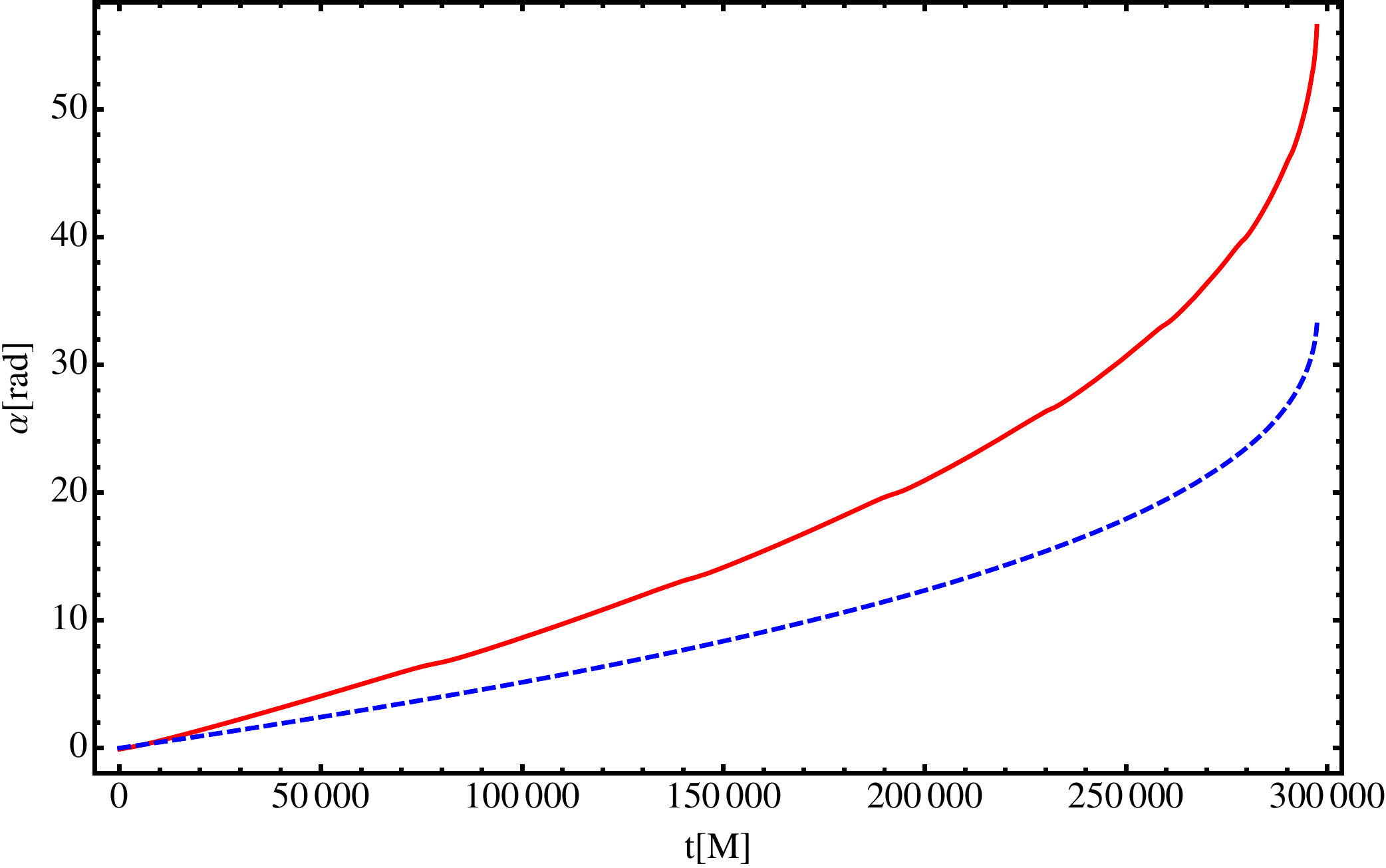}
\includegraphics[width=0.45\textwidth]{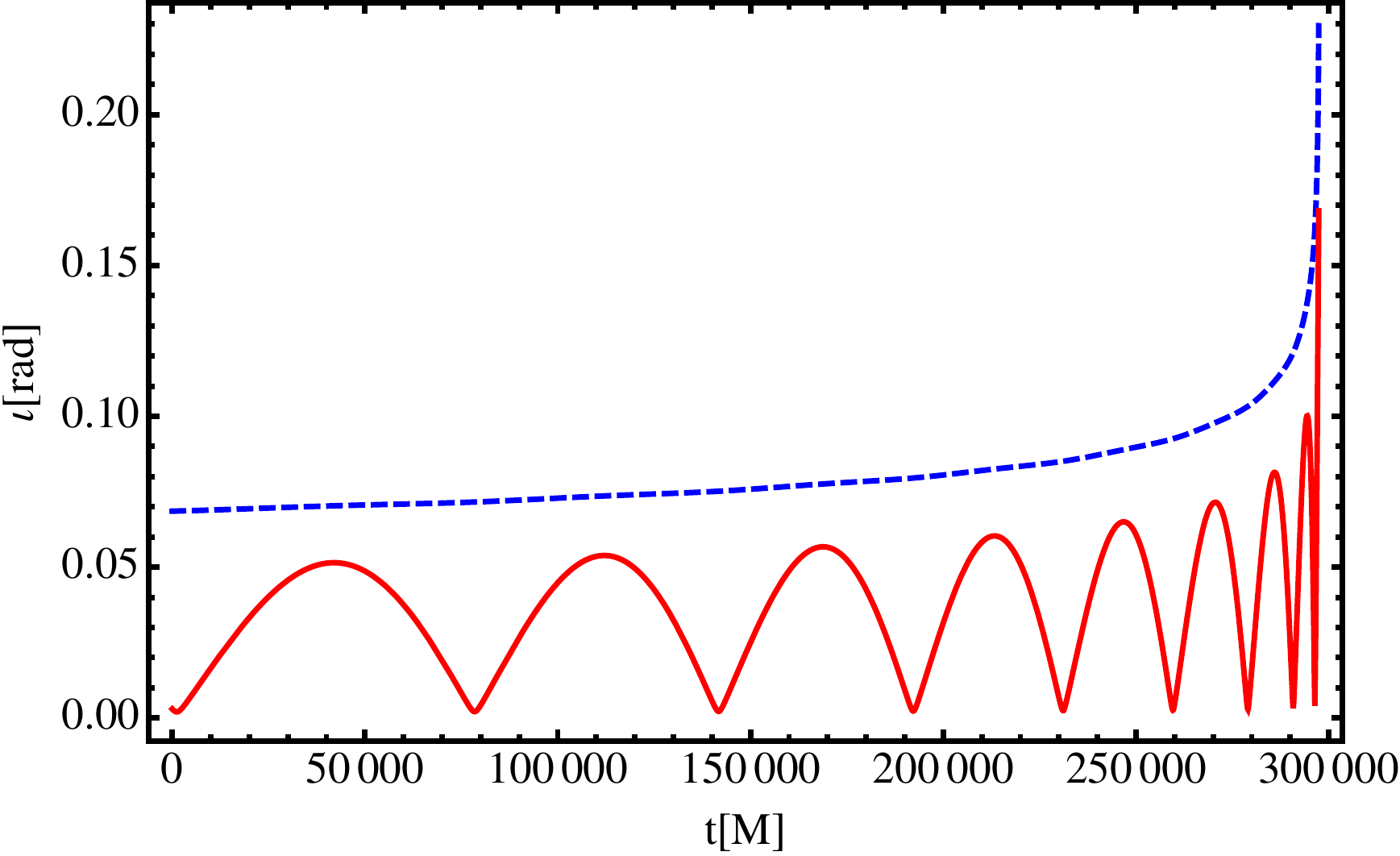}
\caption{The
top panel shows $\alpha(t)$ for the case
$\{q=3,\vec{\chi}_1=(0.38,0.319,-0.079)$,
$\vec{\chi}_2=(-0.036,-0.036,-0.012)\}$ (top solid, red curve) and the corresponding
configuration using $\chi_p$ given by $\{q=3,\vec{\chi}_1=(0.,0.,-0.079)$,
$\vec{\chi}_2=(0.143,0.,-0.012)\}$ (bottom dashed, blue curve); the bottom panel
compares the evolution of the opening angle of the precession cone $\iota(t)$.
Both graphs highlight that in this case $\chi_p$ does not capture the precession
of the system correctly.}
\label{fig:AngleCompSmall}
\end{center}
\end{figure}

One is when the precession is dominated by the in-plane spin of the smaller
black hole. In these cases $\chi_p$ again reproduces the correct contribution
to the precession equation~(\ref{eq:Ldot2}), but the initial value of
$\iota$ may differ more substantially from the correct value. In particular,
$\iota$ is typically small now, and the oscillations in the parallel
and perpendicular spin components (shown in Figs.~\ref{fig:Spara} and
\ref{fig:Sortho}) are now comparable to their mean values. An example is shown
in Fig.~\ref{fig:AngleCompSmall}. We will see in Sec.~\ref{sec:results} that the 
waveforms nonetheless agree well in most cases, and for a wide range of 
binary orientations and \GW polarisations. 

By solving $S_p-A_1S_{1\perp}=0$ for each mass ratio one can define the 
minimal in-plane spin on the larger
black hole as a function of $\chi_{1\perp}$ (the in-plane component of the
smaller black hole) such that the precession is dominated by
$\chi_{1\perp}$. 
For mass ratio $q=3$ and a maximal in-plane spin of
$\chi_{1\perp}=1$, any in-plane spin $\chi_{2\perp}\leq 0.289$
yields a system that is precession-dominated by the smaller black hole; for
$q=10$ this value drops to $\chi_{2\perp}\leq 0.079$, showing that the fraction of 
binaries that are precession-dominated by the smaller black hole decreases with
increasing mass ratio.

\begin{figure}
\begin{center}
\includegraphics[width=75mm]{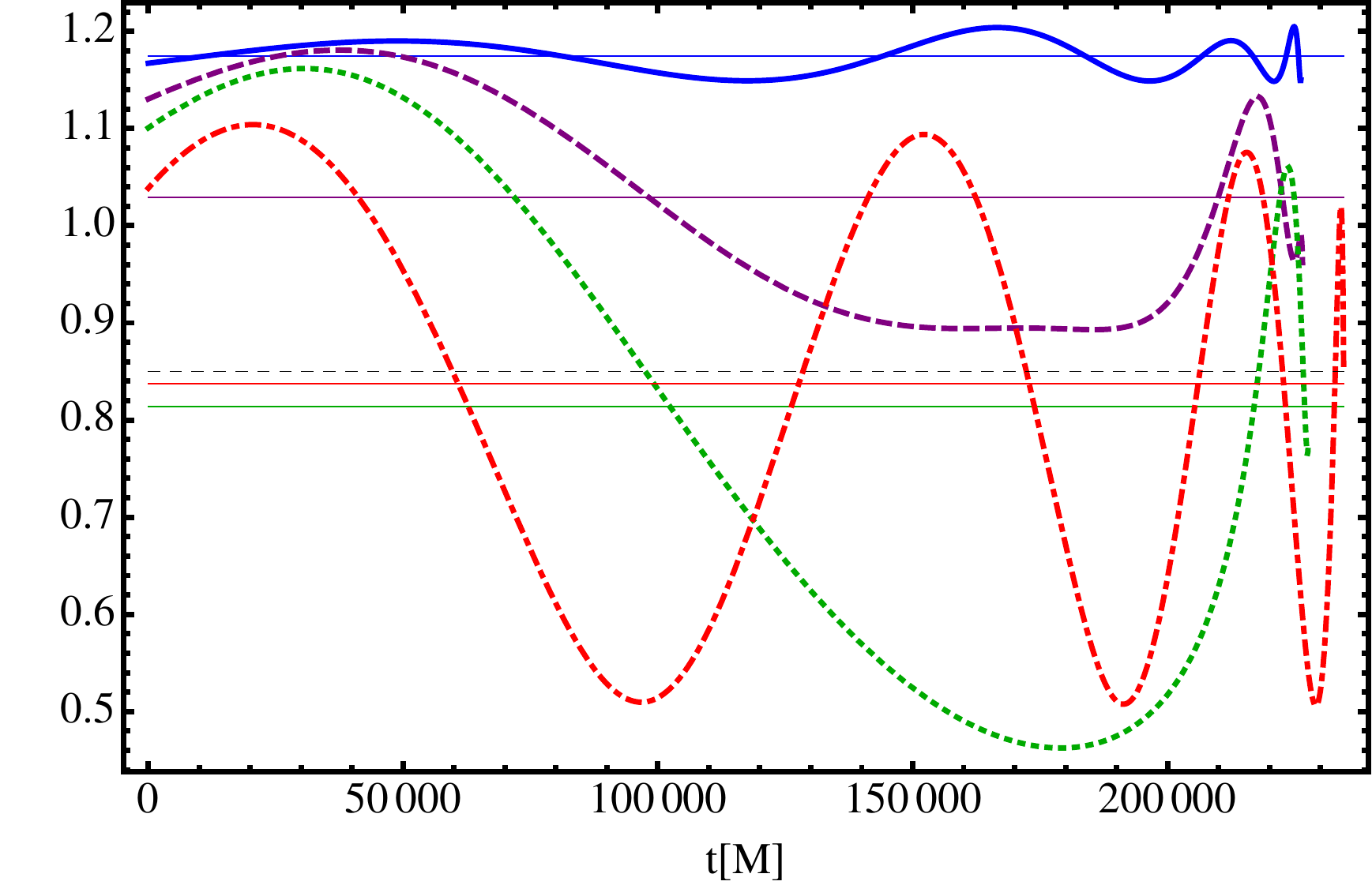} 
\caption{The panel shows the precession term $||A_1
\vec{S}_{1\perp}+A_2 \vec{S}_{2\perp}||/(A_2 m_2^2)$ and its mean
as a function of time for four different mass ratios $q$: equal-mass (blue, solid), $q=1.1$ (purple, dashed), $q=1.2$ (green, dotted) and $q=1.5$ (red, dot-dashed). The mean value for $q=1$ is $1.175$. For this spin configuration, however,
Eq.(\ref{eq:chip}) yields $\chi_p=0.85$ (indicated by the dashed horizontal line). We see that at a small mass ratio of
$q=1.2$, $\chi_p$ is already a good estimator of the average precession.}
\label{fig:ChiPq}
\end{center}
\end{figure}

The second group of configurations where $\chi_p$ will not adequately
approximate the precession dynamics are those where there is little or no
relative rotation of the in-plane spins in the orbital plane. This occurs when
both constituent masses are (almost) equal, i.e., $q \simeq 1$. Then the spins
remain approximately locked  and the averaging that motivates $\chi_p$ no longer
applies. The appropriate choice of in-plane spin magnitude in these cases would
be the \emph{sum} of the two in-plane spin vectors, which remains roughly
constant~\cite{Apostolatos:1994mx}, and so $\chi_p$ tends to underestimate the
in-plane spin contribution. The precession term for varying mass ratio is
illustrated in Fig.~\ref{fig:ChiPq}. We see that, as expected, for the
equal-mass case $\chi_p$ underestimates the average precession of the system. We
see, however, that already at mass ratio $q=1.2$, $\chi_p$ is a good estimator
of the precession even for mass ratios close to equal-mass.

So far, we have explored the phenomenology of a single spin parameter $\chi_p$
to estimate the average precession in a generic system and saw good agreement
when considering precession-related geometric quantities like the precession
angles. However, keeping our goal of modelling precessing waveforms with a
smaller set of physical parameters in mind, we need to investigate and quantify
the agreement between fully generic waveforms and their parameter-reduced
counterparts. This will be the goal of the subsequent sections.

\section{Accuracy of the precession spin approximation: methodology}
\label{sec:pn}
We now assess the quality of our precession parameterisation for \PN inspiral
waveforms. Our approach is to compare a large number of generic inspiral
waveforms at three mass ratios, $q = 1,3,10$, to a family of corresponding
reduced-parameter waveforms where the initial in-plane spin components are
replaced by $\chi_p$ applied to the larger black hole. We are interested only in
the effectiveness of $\chi_p$ to approximate the precession, and so use the 
same values for the masses and
initial values of the parallel spin components $\chi_{i\parallel}$. 

\subsection{Reduced-parameter waveforms}
\label{sec:reduced-waveforms}

We compare a given binary configuration with a full set of physical parameters
with a corresponding configuration with a reduced set of physical parameters,
defined by the mapping of the dimensionless spins as follows:
\begin{align}
\label{eq:map}
    (\chi_{1x},\chi_{1y},\chi_{1z}) &\mapsto (0,0,\chi_{1z}),    \nonumber \\ 
   (\chi_{2x},\chi_{2y},\chi_{2z}) &\mapsto (\chi_p,0,\chi_{2z}),
\end{align}
where we have defined the spins with respect to $\hat{L} \equiv \hat{z}$ in a
Cartesian coordinate system. Hence, the reduced model parameters are: $q,
\chi_{1||}, \chi_{2||}$ and $\chi_\mathrm{p}$. 

This does not define a bijective map: various combinations of different physical
spins $\vec{S}_1, \vec{S}_2$ can yield the same set of $\{\chi_{1||},\chi_{2||},
\chi_\mathrm{p}\}$ despite being physically completely different configurations.
Therefore, all configurations for one set of model parameters
$\{q,\chi_{1||},\chi_{2||},\chi_p\}$ do not define a single configuration but an
\emph{approximate equivalence class} of precessing systems, i.e., various
generic configurations map to the same point in the manifold of
reduced-parameter configurations. If we are correct in assuming that all of
these configurations agree well with each other, then this has implications for
\GW observations: we will be able to more easily measure the combination
$\chi_p$ than the individual in-plane black-hole spins. 

In order to assess whether this approximation indeed holds, we compute waveforms
by integrating the set of \PN equations given in Appendix~\ref{sec:wfgen}. We
then compute matches between waveform strains of the generic
configuration and its corresponding reduced-parameter configuration for various
binary inclinations $\theta$ and \GW polarisation angles $\psi$. Henceforth,
motivated by the terminology of \GW searches, we will refer to the
full-parameter configuration as \emph{signal} and to the reduced-parameter one
as \emph{model}. 

\subsection{Generic match} \label{sec:gen_match}
The agreement between two waveforms is commonly quantified by the noise-weighted
inner product between the two signals~\cite{lrr-2012-4}. In the case of the
real-valued detector response, $\hr^S(t), \hr^M(t) \in \mathbb R$ (where the
superscripts distinguish signal and model), the match is commonly defined as
\begin{eqnarray}
 \inner{\hr^S}{\hr^M} &=& 2 \int_{-\infty}^\infty \frac{\fhr^S(f) \,
\fhr^{M\ast}(f)}{S_n(\vert f \vert)} d\!f  \label{eq:match1} \\ 
&=& 4 \Re \int_0^\infty \frac{\fhr^S(f) \,
\fhr^{M\ast}(f)}{S_n(\vert f \vert)} d\!f . \label{eq:match2}
\end{eqnarray}
Here, $S_n$ is the noise spectral density of the detector, $\tilde x$ denotes
the Fourier transform of $x$, and $x^\ast$ is the complex conjugate of $x$. 

Note that the conversion from (\ref{eq:match1}) to (\ref{eq:match2}) relies on
$\fhr(-f) = \fhr^\ast(f)$ which is always true for real-valued signals. Here we
find it more convenient, however, to work directly with a commonly used
\emph{complex} waveform strain that combines both waveform polarisations, 
\begin{equation}
 h = h_+ - i \, h_\times.
\end{equation}

With the introduction of a polarisation angle $\psi$, we can relate both
waveform representations to each other via
\begin{eqnarray}
 \hr(t) &=& \cos(2\psi) \, h_+(t) + \sin(2\psi) \, h_\times(t) \\
     &=& \Re\left[ h(t) \; e^{ i2\psi} \right]. \label{eq:complex_polarisation}
\end{eqnarray}
Note that our definitions of $h_+$ and $h_\times$  differ slightly 
from similar expressions in the literature (see, e.g., Eq. (55) in
\cite{lrr-2009-2}) in the respect that we leave an overall factor that depends
on the orientation between detector and source as part of the definitions of
$h_+$ and $h_\times$, while $\psi$ explicitly governs a relative rotation in
the detector plane.  

Our goal is to calculate the inner product between signal and
model and optimize it over the model polarisation angle and a relative time
shift in an efficient way. We find a convenient formulation of the inner
product in terms of the complex strains by inserting
(\ref{eq:complex_polarisation})
into (\ref{eq:match1}), which finally yields
\begin{eqnarray}
\inner{\hr^S}{\hr^M} = \mathrm{Re}
\int_{-\infty}^\infty \frac{\tilde h^S(f) \,
\tilde h^{M\ast}(f)}{S_n(\vert f \vert)} e^{2i(\psi_S-\psi_M)} d\!f \nonumber
\\*
 + \mathrm{Re} \int_{-\infty}^\infty \frac{\tilde h^S(f) \,
\tilde h^{M}(-f)}{S_n(\vert f \vert)}e^{2i(\psi_S+\psi_M)}d\!f.
\label{eq:fullmatch}
\end{eqnarray}
The details
of the derivation are given in Appendix~\ref{sec:precmatch}, where we also
provide explicit expressions to optimize over $\psi_M$ (for a given signal
polarisation $\psi_S$)
analytically. 

Note that nonprecessing signals under the adiabatic assumption have all
information contained on one side of the frequency spectrum, hence the second
term in (\ref{eq:fullmatch}) vanishes. Here, however, we do not make
this assumption about the (precessing) signals; in fact, for 
orientations where the \GW strain is not dominated by only one mode and
precession features become important, we have to take into account
both contributions in (\ref{eq:fullmatch}) to obtain the correct inner
product.

The results presented in the next section are all formulated in terms of the
\emph{match} $\mathscr M$, which we define as the inner product
(\ref{eq:fullmatch}) normalised by both signal powers and optimised over a
relative time shift, the polarisation angle $\psi_M$ of the model, and the
azimuthal angle $\varphi_M$ in the spin-weighted spherical harmonics of the model (see Eq.~(\ref{eq:strain}) for more
 details). For
details of the algorithm, we refer once again to Appendix~\ref{sec:precmatch}.
Alternative approaches to similar problems have been introduced before in
\cite{Damour:1997ub} and were extended in
\cite{Vaishnav:2007nm, McWilliams:2010eq}, but these
relied on the contruction of an orthogonal basis and expressed the results in
terms of matches that were maximised or minimised over $\psi_S$. Here, however,
we
prefer to directly use the information from the complex \GW strains
across the entire frequency spectrum as this is what we obtain from the \PN
integration.

Matches (very) close to unity indicate an accurate approximation of the
full signal, while any deviation from unity quantifies the degree of
disagreement between model and signal. There are various application-dependent
thresholds one could consider for $\mathscr M$, some being based on the
distinguishability between model and signal, others translating mismatches to a
loss in sensitive volume~\cite{Lindblom:2008cm}. For simplicity, we will
use $\mathscr{M} =0.965$ as a reference value, as this number is
frequently used in the \GW literature to mark the 10\% loss in sensitive volume.
We remark, however, that we are not explicitly addressing the question of detecting
the signal with our proposed model. We deliberately refrain from optimising the
match over all intrinsic source parameters (which would be a meaningful strategy
to quantify the detection efficiency), instead we quantify the agreement for
fixed source parameters (with the exceptions pointed out above) because we are
predominantly interested in whether our reduced-parameter model introduced in
Secs.~\ref{sec:chip} and \ref{sec:reduced-waveforms} faithfully represents the
full-parameter signals. 

In the following, we quantify the agreement between the $(\ell =2)$-waveform strain
of the signal
\begin{equation} \label{eq:strain}
h(t;\theta,\varphi)=\sum_{m=-2}^{2}h_{2m}(t) Y^{-2}_{2m}(\theta,\varphi),
\end{equation}
by exploring the match
$\mathscr{M}$ against the model as a function of the binary inclination $\theta$ and the signal
polarisation $\psi_S$ for a total binary mass of $M=12M_\odot$ with a GW
starting
frequency of $20\mathrm{Hz}$ and a cutoff frequency of $366\mathrm{Hz}$. We use
the early aLIGO noise curve~\cite{Shoemaker:2010}.
 
\section{Accuracy of simplified precessing inspiral waveforms: results}
\label{sec:results}

In the following we perform two classes of tests of our reduced-parameter model.
We first test the $\chi_p$ parameterisation on a selected set of configurations
where one or both black holes have extremal spins: we vary the
relative orientation of the in-plane spins of the signal configuration
(Sec.~\ref{sec:orientation}), the magnitude of one of the in-plane components
(Sec.~\ref{sec:magnitude}), and assess the influence of the parallel spin components
(Sec.~\ref{sec:parallel}). Having tested the parameterisation in what we
consider to be extreme cases, we then analyse in Sec.~\ref{sec:stats} a large
sample of configurations with three different mass ratios, $q = 1,3, 10$, with
randomly chosen spins magnitudes and orientations, and a selection of
binary orientations and polarisations.

We emphasise that the faithfulness we calculate is the lower bound for the
model's detection effectualness as no optimisations over physical parameters are
performed; if we were to optimise over physical parameters as done in a \GW
search, the resulting fitting factor would by definition be larger (or the
same). The results show very strong evidence in favour of the reduced
parameterisation to capture the dominant precession effects. 

\subsection{Selected test cases}
To test the effectiveness of the reduced
parameterisation, we first explore double-spin binaries with either one or two 
maximally spinning black holes. In the following, we analyse
various properties of these particular configurations for the mass ratio $q=3$. 

\subsubsection{Relative in-plane spin orientation}
\label{sec:orientation}
The first investigation concerns the influence of the \emph{relative orientation
of the spins in the plane}. Apart from the spin-spin terms in the
\PN evolution equations, the relative orientation of the spins 
has no impact on the waveform at quadrupole order. In that sense, we are now
testing the influence of the spin-spin terms. 

\begin{figure*}
\begin{center}
\includegraphics[width=75mm]{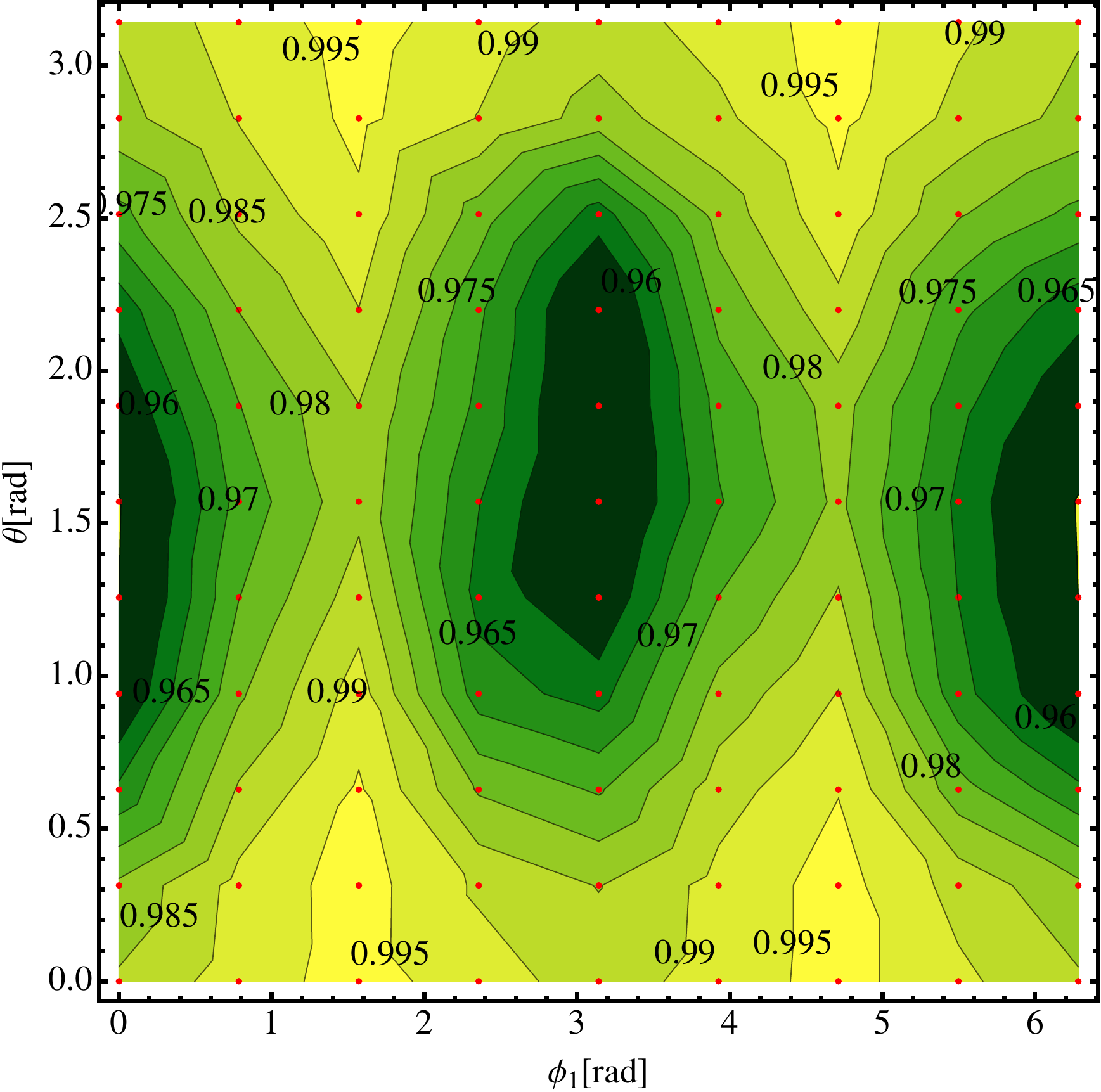}
\quad
\includegraphics[width=75mm]{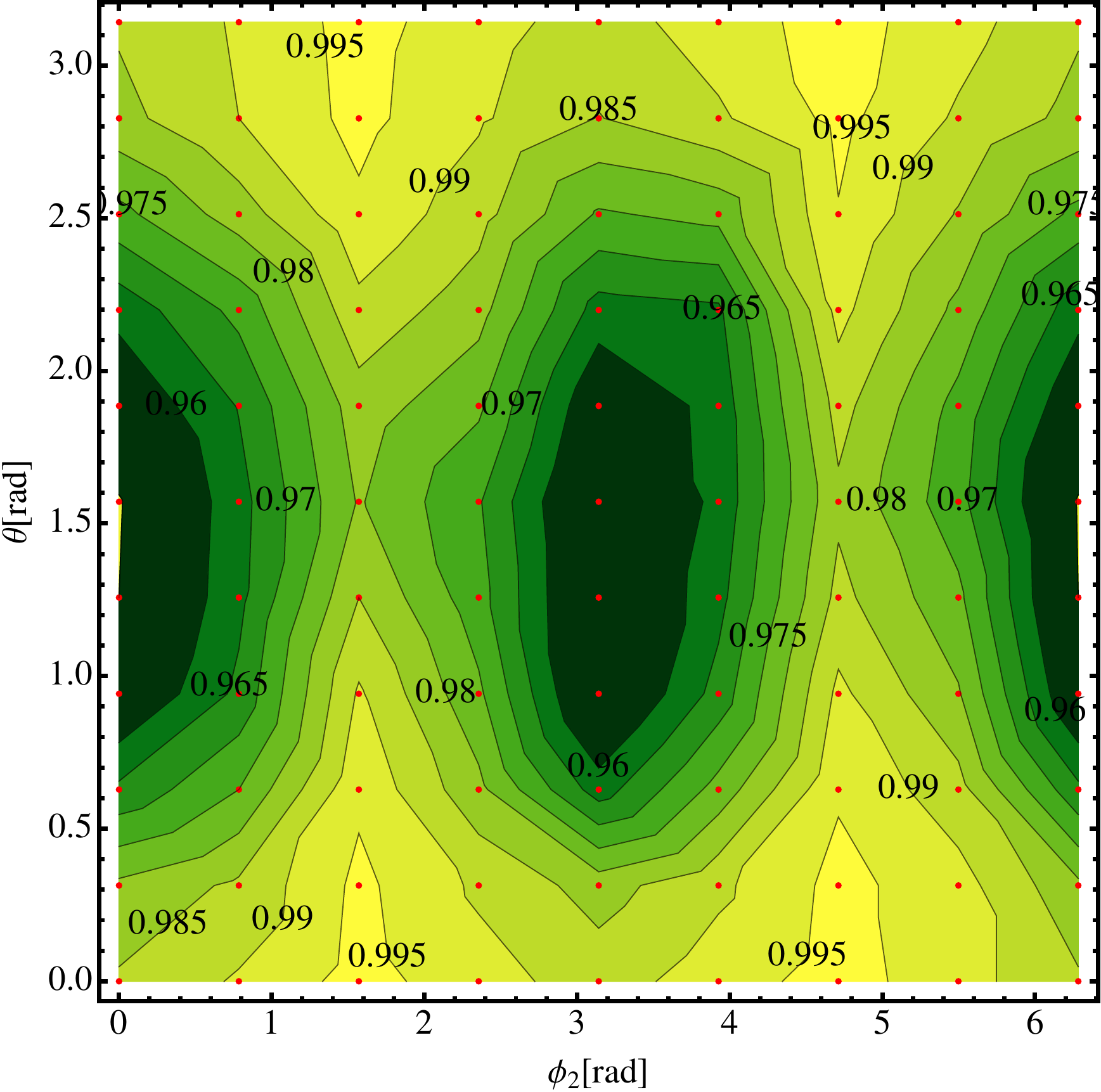}
\caption{The left panel shows the match contours for the extremal case with
$\vec{\chi}_2 = (1,0,0)$ and varying in-plane orientation of $\vec{\chi}_1$, while 
the right panel shows the contour for $\vec{\chi}_1 = (1,0,0)$ and varying orientation of 
$\vec{\chi}_2$ as a function of the binary
orientation $\theta$. The red dots mark the actual points at which the matches
are evaluated.}
\label{fig:RelativeRot1}
\end{center}
\end{figure*}

We first fix $\vec{\chi}_2\equiv ({1,0,0})$ and vary the orientation of
$\vec{\chi}_1=(\cos\phi_1, \sin \phi_1,0)$ with $\phi_1 \in [0, 2\pi]$ and
$\Delta \phi_1 =45^\circ$. We then interchange the roles of $\vec{\chi}_1$ and
$\vec{\chi}_2$ and vary $\phi_2$ in the same interval. To quantify the agreement
between each rotated generic waveform and our model waveform that remains
unaffected by these rotations, we compute the match between these two,
respectively. We choose a set of different binary orientations $\theta \in
[0,\pi]$ with $\Delta \theta = \pi/10$, but keep the signal polarisation fixed
for a polarisation angle $\psi_S=0$ and set the azimuthal orientation of the
signal to
$\varphi_S=0$. We optimise the
match over the template polarisation, a time shift and the angle $\varphi_M$ in
the spin-weighted spherical harmonics of the template strain.

The results are illustrated in Fig.~\ref{fig:RelativeRot1}. In both cases we
obtain very high matches but observe 1) a mild dependence on the relative
orientation in the plane and 2) a strong dependence on the binary's orientation
$\theta$. The minimal match is $\mathscr{M}_{\mathrm{min}}=0.95$ in both cases.
We find that the lowest matches are clustered around ``edge-on'' orientations of
$\theta = \pi/2$. 

The pattern of low matches in Fig.~\ref{fig:RelativeRot1} can be explained by
considering the \PN evolution equation~(\ref{eq:rdot}). The spin-spin
($\vec{S}_1 \cdot \vec{S}_2$) contribution vanishes completely in the
reduced-parameter system which in this case only has one non-vanishing spin.
However, the full system does have a spin-spin contribution, and this is
maximised at the beginning of the evolution when $\phi = n\pi$. In these cases
the inspiral rate, and therefore the \GW phase evolution, will
differ during the early part of the evolution. The evolution of all orbital
components is slower at earlier times, and so the level of agreement in the
early phase of the evolution has the strongest influence on the overall
agreement of the two final waveforms. This explains why the matches are lowest
around $\phi = n\pi$. We emphasise, however, that this is purely based on the
fact that we indicate the phase when the signal enters the detector band. If we
were to show spin angles at different times or frequencies, the pattern
in Fig.~\ref{fig:RelativeRot1} would shift. The location of the poor-match
regions with respect to an arbitrary $\phi$ has no physical significance.

\subsubsection{Varying the in-plane spin magnitude}
\label{sec:magnitude}
\begin{figure*}
\begin{center}
\includegraphics[width=75mm]{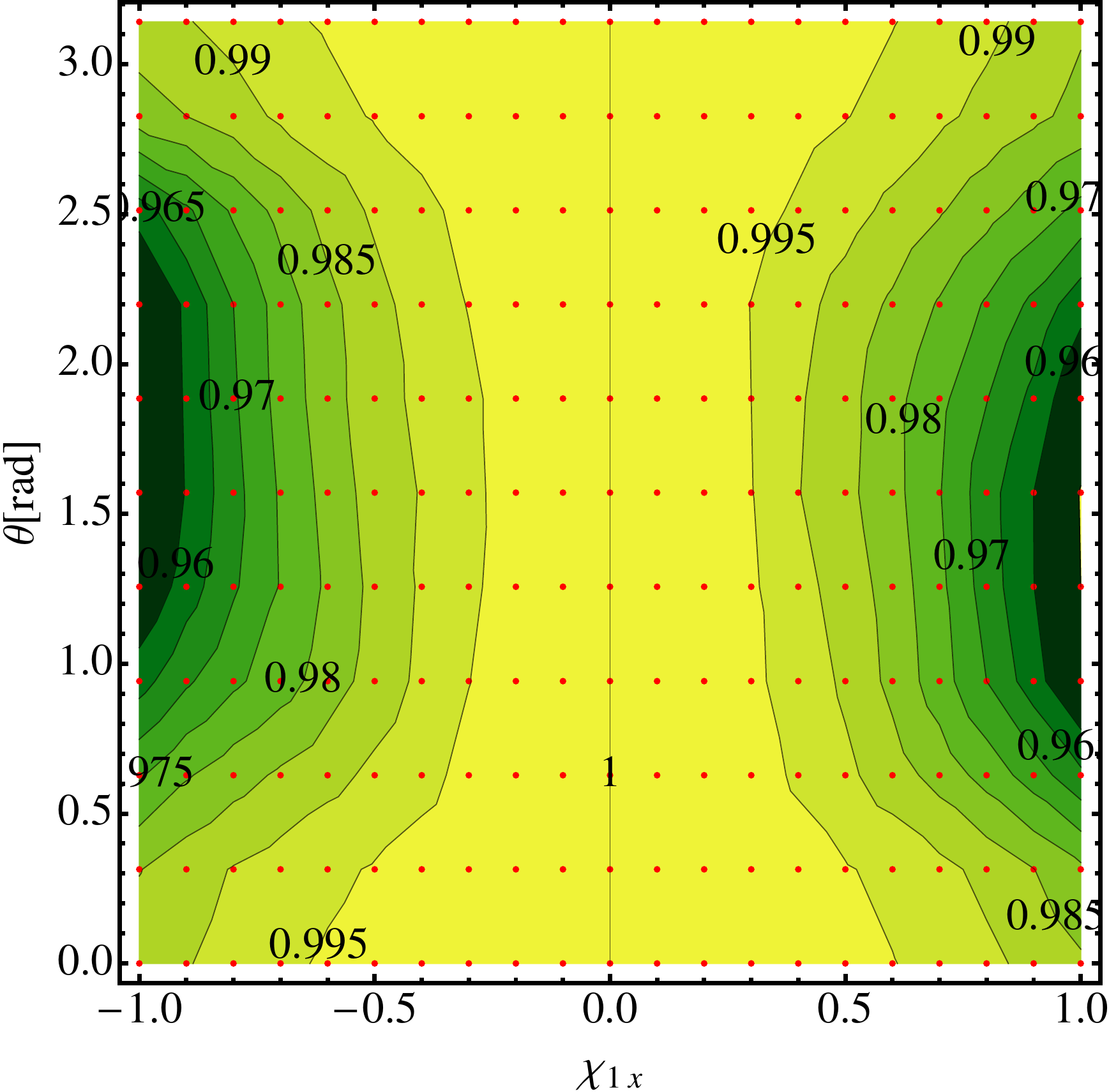}
\includegraphics[width=75mm]{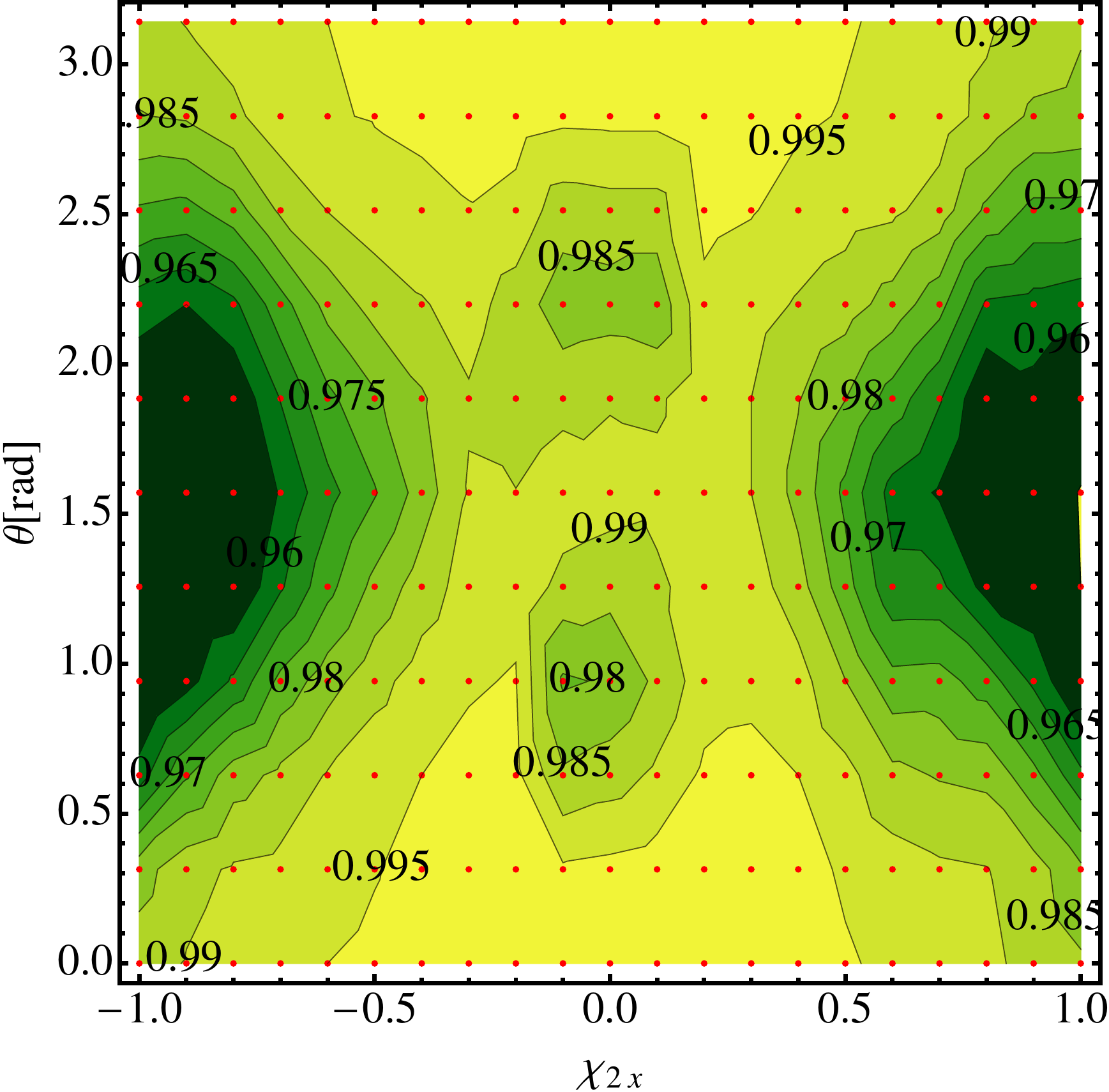}
\caption{The left panel shows the match for $\vec{\chi}_1=(\chi_{1x},0,0)$ and
$\vec{\chi}_2=(1,0,0)$ against the appropriate reduced-parameter waveforms as a
function of the binary orientation; the right panel shows the match for
$\vec{\chi}_1=(1,0,0)$ and $\vec{\chi}_2=(\chi_{2x},0,0)$ against the
appropriate reduced-parameter template waveforms. The red dots mark the actual
configurations used to obtain the contours.}
\label{fig:VaryMag1}
\end{center}
\end{figure*}

In this section we investigate the influence of the in-plane spin magnitude. We fix
the relative spin orientation to $\phi_1-\phi_2=0$ in this study as we have seen
earlier that initially parallel in-plane spins yield the lowest matches for
certain orientations. As before, the signal polarisation is fixed such that
$\psi_S=0$ and we choose $\varphi_S=0$; we compute the match for various binary
orientations. Firstly, we let $\vec{\chi}_2=(1,0,0)$ and vary the magnitude of
the spin on the smaller black hole such that $\vec{\chi}_1=(\chi_{1x},0,0)$. We
then exchange the role of the two black holes and vary
$\vec{\chi}_2=(\chi_{2x},0,0)$ and set $\chi_{1x} = 1$. The contours for the
matches as a function of the in-plane spin magnitude of one of the holes and the
binary inclination $\theta$ is shown in Fig.~\ref{fig:VaryMag1}.

We find that the magnitude of the in-plane spin of the smaller black hole is
negligible up to $\vert \chi_{1x} \vert \simeq 0.8$, and for $\vert \chi_{2x}
\vert \simeq 0.7$. The lowest matches are recovered for maximal in-plane spins
on both black holes, which is consistent with the results regarding the relative
orientation. Again, we can attribute decreasing matches to the growing influence
of the spin-spin coupling term that is proportional to the individual spin
magnitudes; our simplified model discards parts of these terms completely. We also
observe additional structures in the match contours when $\vec{\chi}_1$ is fixed
and the in-plane spin magnitude of $\vec{\chi}_2$ is varied, in particular for
$\vert \chi_{2x} \vert \simeq 0$. 

\subsubsection{The influence of parallel spins}
\label{sec:parallel}
\begin{figure*}
\includegraphics[width=0.3\textwidth]{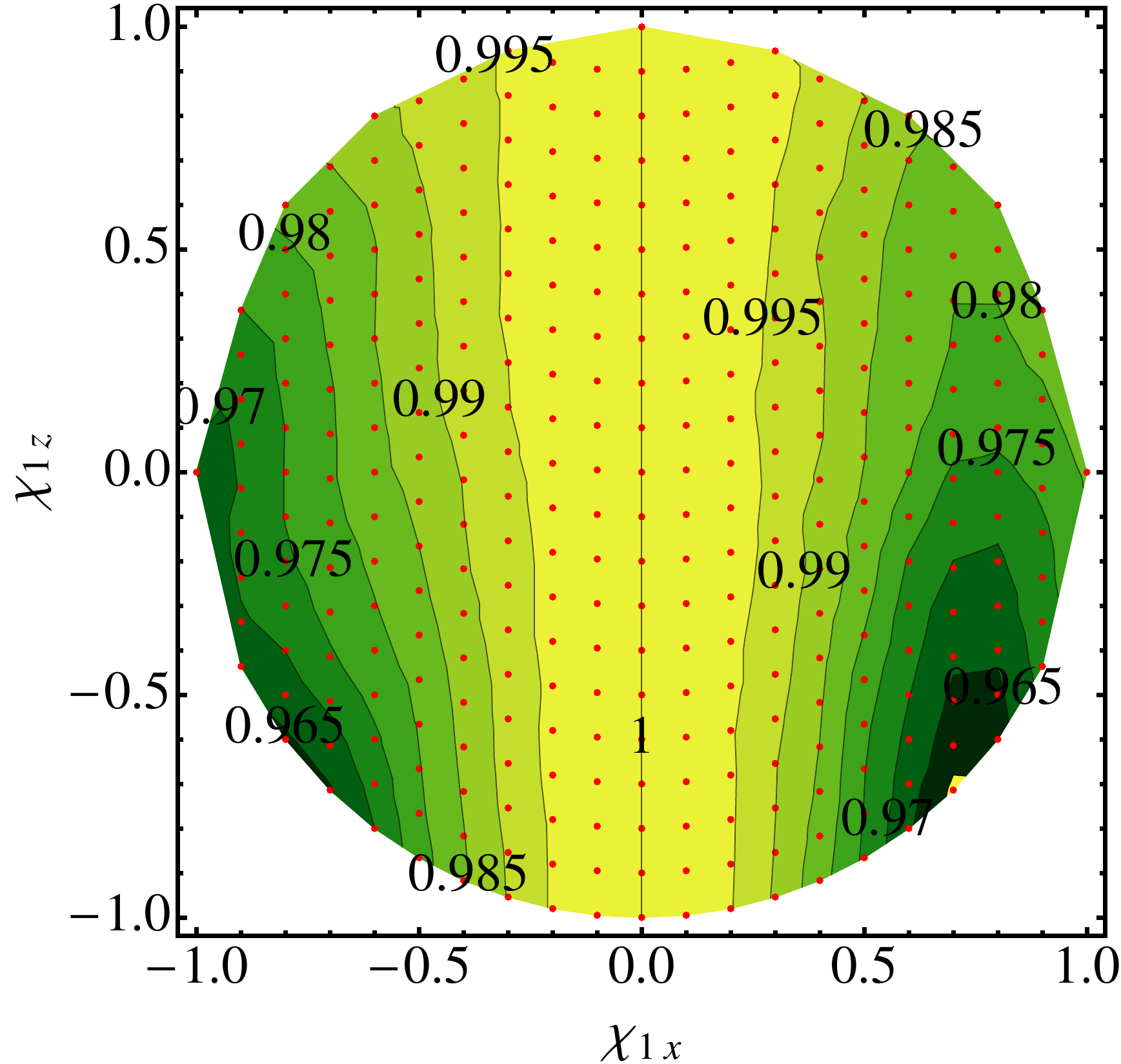}
\quad
\includegraphics[width=0.3\textwidth]{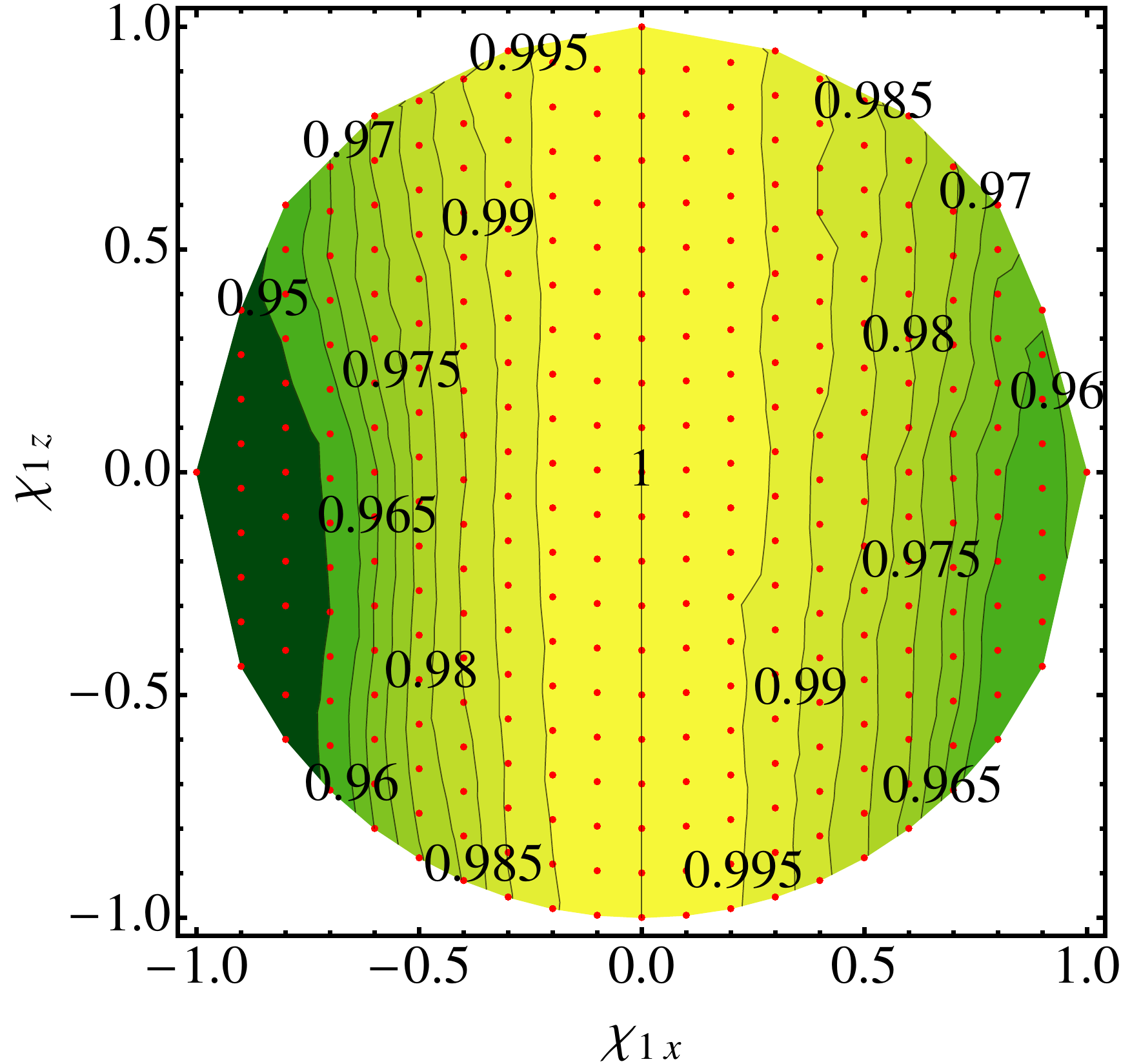}
\quad
\includegraphics[width=0.3\textwidth]{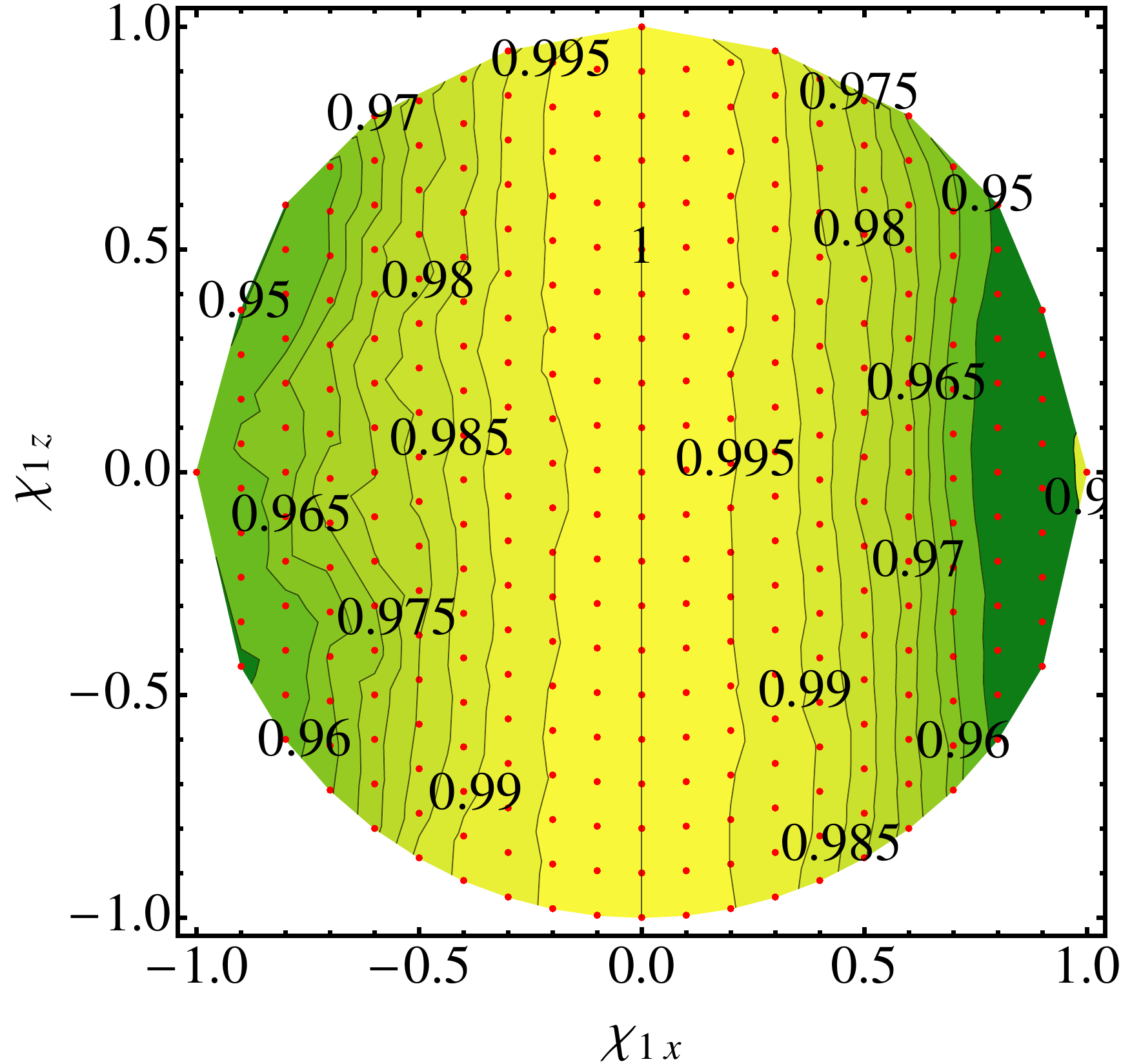}
\caption{The panel shows the match contours for three different binary inclinations ($0^\circ,36^\circ$ and $90^\circ$) for the configurations where $\vec{\chi}_2=(0.8,0,-0.6)$ and $\vec{\chi}_1=(\chi_{1x},0,\chi_{1z})$. Each red dot represents one particular choice of $(\vec{\chi}_1,\vec{\chi}_2)$. We find that the matches drop with increasing value of $\chi_{1x}$ and decrease overall with increasing inclination $\theta$.}
\label{fig:ParallelSpins}
\end{figure*}

In the cases we have considered so far, we have set the parallel components of
the spins initially to zero so that they exhibit only small oscillations around
zero throughout the inspiral. As described earlier, the precessional dynamics
decouples approximately from the inspiral dynamics, and therefore in these cases
we have studied precession effects with minimal spin influence on the inspiral. 

We now introduce non-zero parallel spin components and therefore study our
reduced parameterization for different inspiral rates. We consider the following
configuration: the spin on the larger black hole is fixed and set to
$\vec{\chi}_2=(0.8,0,-0.6)$ ($\chi_2=1$); we now vary the spin of the smaller
black hole $\vec{\chi}_1=(\chi_{1x},0,\chi_{1z})$. The mass ratio is again $q =3$. 

The results for three binary inclinations $\theta=0^\circ, 36^\circ, 90^\circ$ 
and signal polarisation $\psi_S=0$ are shown in Fig.~\ref{fig:ParallelSpins}. 
The lowest match we
obtain is $\mathscr{M}_\mathrm{min}=0.826$ for the configuration with
$\vec{\chi}_1=(-1,0,0)$. Following Eq.~(\ref{eq:map}), the parallel components
of the model waveform are the same as in the generic signal. Keeping this in
mind, Fig.~\ref{fig:ParallelSpins} can be interpreted as follows: if
$\chi_{1\perp}=0$, then the reduced system exactly corresponds to the generic
system and we therefore obtain matches $\mathscr{M}=1$. For $\chi_{1z}=0$ we see
a decreasing agreement with increasing $|\chi_{1\perp}|$ due to the neglect of
the in-plane contribution to the spin-spin coupling. In between these extremes
we see nearly vertical contours indicating that the mismatch is indeed dominated
by the neglect of $(\vec{S}_{1\perp} \cdot \vec{S}_{2\perp})$ and rather independent of
the parallel spin components as these are preserved in the particular mapping
and PN treatment we use.

\subsection{Statistical analysis: a random sample of precessing configurations}
\label{sec:stats}

\begin{figure*}
\includegraphics[width=80mm]{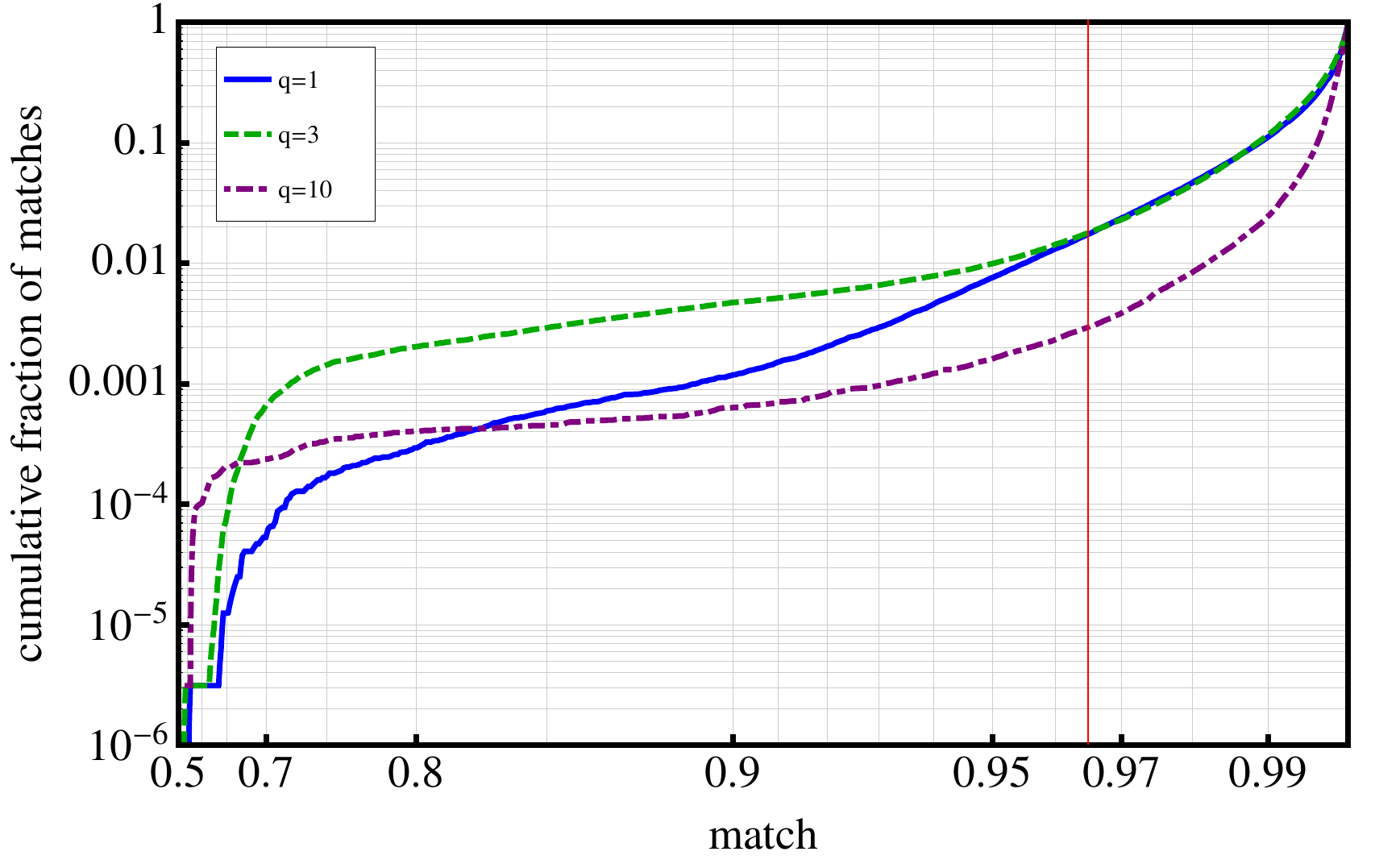}
\includegraphics[width=80mm]{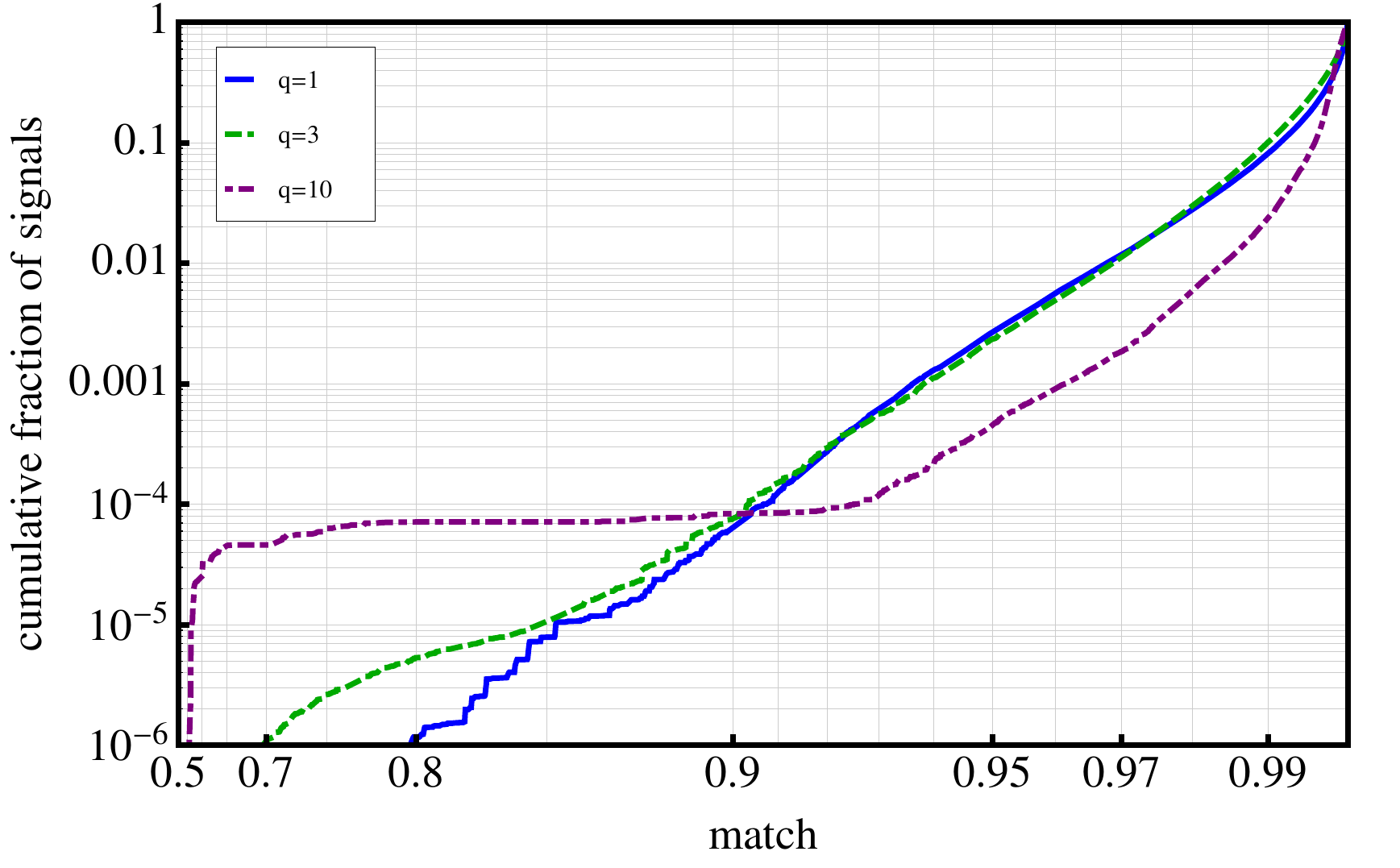}
\caption{The left panel shows the cumulative distribution function for all
matches for the mass ratio $q=1$ (blue, solid), $q = 3$ (green, dashed) and
$q=10$ (purple, dot-dashed). The red vertical line indicates a match of
$\mathscr{M}=0.965$. In the right panel the distribution is weighted according
to the signal strength to represent the fraction of actually detectable
signals.} 
\label{fig:q1CDF}
\end{figure*}

Previously, we have analysed a handful of test cases, which allowed us to
extract trends along several directions in the configuration space. Further, we
were able to quantify the influence of the in-plane spin-spin coupling, which is
neglected in our approximation. In order to assess the goodness of the
reduced-parameter model across the precessing binary parameter space, a
significant sample of all possible configurations needs to be
analysed at various mass ratios. We therefore construct 10,000 random binary
spin configurations with uniform sampling in the dimensionless spin magnitudes
$\chi_{1,2} \in [0,1]$ and the spin azimuth angles $\phi_{1,2} \in [0,2\pi]$, for
mass ratios $q = 1, 3$ and 10. 

We analyse the sample by quantifying the agreement between the
$(\ell=2)$-waveform strain as given in Eq.~(\ref{eq:strain})
for each configuration in the sample with its corresponding
reduced-parameter model $h^M$ by computing the match $\mathscr{M}$. 
As before,
we optimise only over the following subset of extrinsic parameters: the
polarisation $\psi_M$ of the model waveform, the azimuth $\varphi_M$ in the
reduced-parameter \GW strain, as well as a time shift $\Delta t$; we do not
optimise over the physical parameters $m_1$, $m_2$, $\vec{S}_1$ or $\vec{S}_2$.
We repeat this match computation for each configuration for the signal
polarisation angles $\psi_S \in \{0.,\pi/8,\pi/4,3\pi/8 \}$ as well as for the
binary orientations $\theta \in \{ 0,\pi/10, \pi/4,
2\pi/5,\pi/2,3\pi/5,4\pi/5,11\pi/12\}$ with $\varphi_S=0$. This yields 32
individual matches per configuration and a total of 320,000 matches. We repeat
this calculation for various mass ratios but fix the following parameters in the
analysis: the initial separation $r_i=40M$, to obtain sufficiently long inspiral
waveforms in the time domain, which are sampled at intervals of $\Delta t =
10M$. We set the total mass to $M=12M_\odot$. This is an ad hoc choice, but was
made to allow a wide frequency range in the detector's sensitivity band, to
minimise the effects of merger and ringdown and for reasons of computational
cost efficiency. We fix the upper cutoff frequency to be
$\mathrm{Mf}_{\mathrm{ISCO}}=(\pi 6^{3/2})^{-1}$ and use the anticipated early
PSD noise curve for aLIGO~\cite{Shoemaker:2010}.

\subsubsection{General results}

We present the results of our large-scale study in Fig.~\ref{fig:q1CDF}, where
we show the \acp{CDF} for our statistical samples: for each value of the match,
the
figure indicates the fraction of signals that have a match less than that
value. In the left panel, this fraction is simply based on the number of matches
we have calculated. On right panel, however, we estimate the fraction of
\emph{detectable signals} by re-interpolating our
results over a uniform grid in $\cos \theta$ and by assigning a
signal-to-noise-ratio--dependent volume to each signal. By comparison we see
that most signals which are not well captured by our model are unlikely to be
detected (e.g., edge-on signals with pronounced precession effects have a
considerably smaller amplitude than less modulated face-on signals at the same
distance), therefore the right panel of Fig.~\ref{fig:q1CDF} shows generally better
results than the left panel. In order to be conservative and emphasise
the modelling (i.e., amplitude independent) focus of this paper, we shall only
quote numbers obtained from the left panel of Fig.~\ref{fig:q1CDF} below.

For mass ratios $q = 1, 3$ we find that less than $2\%$ of all matches are below
$0.965$, respectively, showing that the precession in the system is faithfully
represented by the effective precession parameter $\chi_p$ for most binary
configurations and orientations. For both mass ratios, more than $88\%$ of all
matches are above $0.99$. We find a difference in the \CDF tails towards low
matches, where the $q=1$ curve is considerably flatter than for $q=3$, which is
rather surprising at first glance. It can be explained by the  error introduced
for unequal-mass cases with very little precession, which are not well captured
by $\chi_p$. For completeness, we remark that the minimum match for $q=1$ is
found at $\mathscr{M}_\mathrm{min}=0.558$ for the following spin configuration:
$\vec \chi_1=(0.14, 0.13, 0.75)$ and $\vec \chi_2=(0.12, 0.22, -0.42)$. For
$q=3$ we find the minimum at $\mathscr{M}_\mathrm{min}=0.532$ for the
configuration $\vec \chi_1=(0.53, -0.04, -0.63)$ and $\vec \chi_2 =(-0.16, 0.18,
0.76)$.

Additionally, we have computed the matches for mass ratio $q=3$ with a random
choice of $\varphi_S$ and obtain a \CDF that shows no significant deviation from
the result when $\varphi_S=0$.

As we have mentioned at the end of Sec.~\ref{sec:chip}, one might not have
expected our $\chi_p$ parameterisation to work accuractly in the equal-mass case
as the two spins are locked and therefore the binary follows the evolution of a
single spin binary with a total spin magnitude
$S=||\vec{S}_1+\vec{S}_2||=\mathrm{const}$. The
appropriate parameter reduction for $q = 1$ configurations might therefore
be to put the \emph{total} spin, $\vec{S} = \vec{S}_1 + \vec{S}_2$ onto
the larger black hole, which would be equivalent to the reduction used in the
Physical Template Family~\cite{Buonanno:2004yd}. However, we find that this
choice has little effect on the results in Fig.~\ref{fig:q1CDF}. This indicates
that in these configurations, the impact of neglecting the spin-spin terms (by
placing all of the spin on one black hole) is comparable to that of making our
$\chi_p$ parameter reduction. As the mass ratio increases, the $\chi_p$
parameterisation becomes more accurate, and the influence of the spin-spin terms
to the phase evolution decreases.

\begin{figure*}
\begin{center}
\includegraphics[width=80mm]{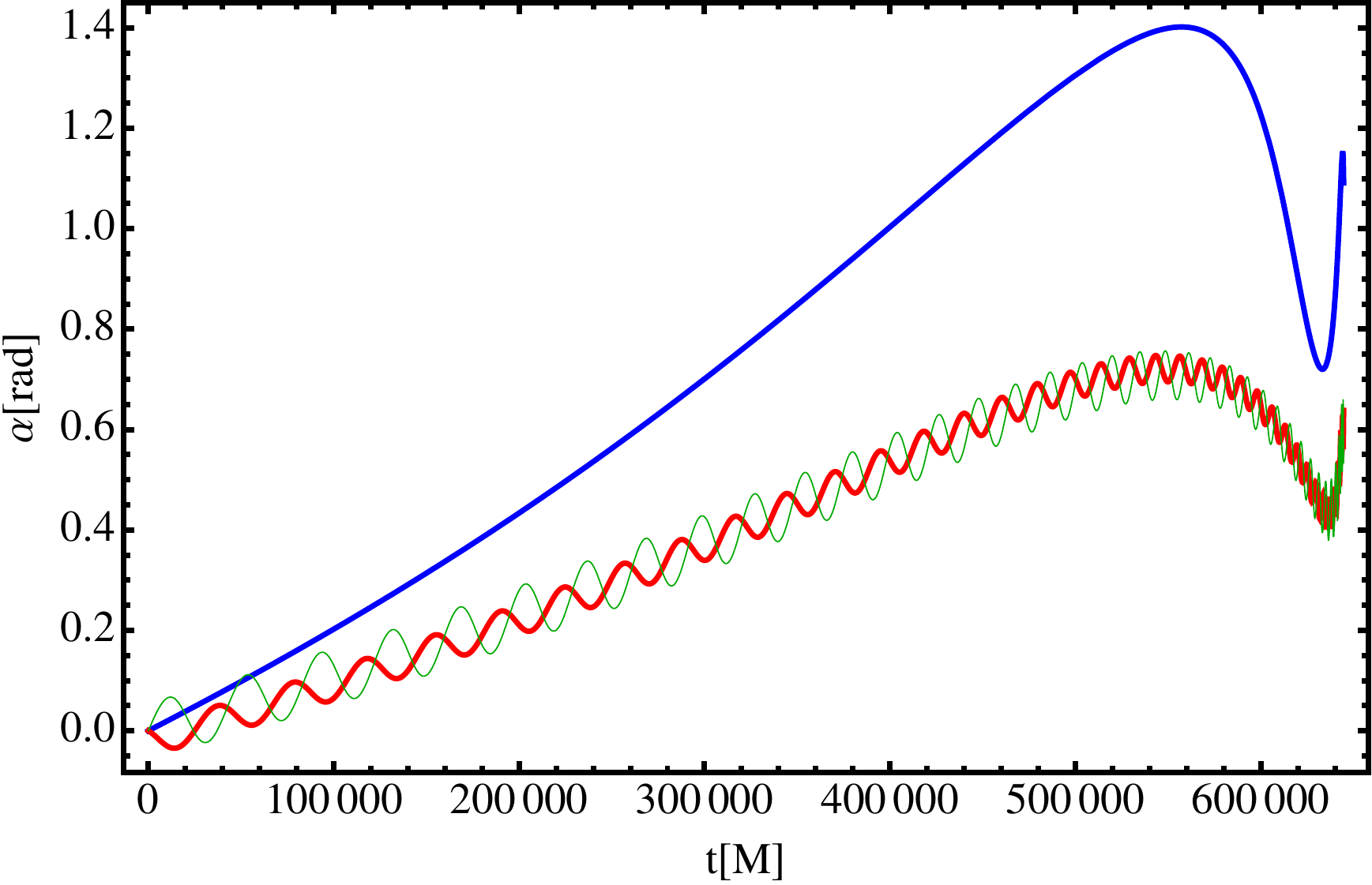}
\quad
\includegraphics[width=80mm]{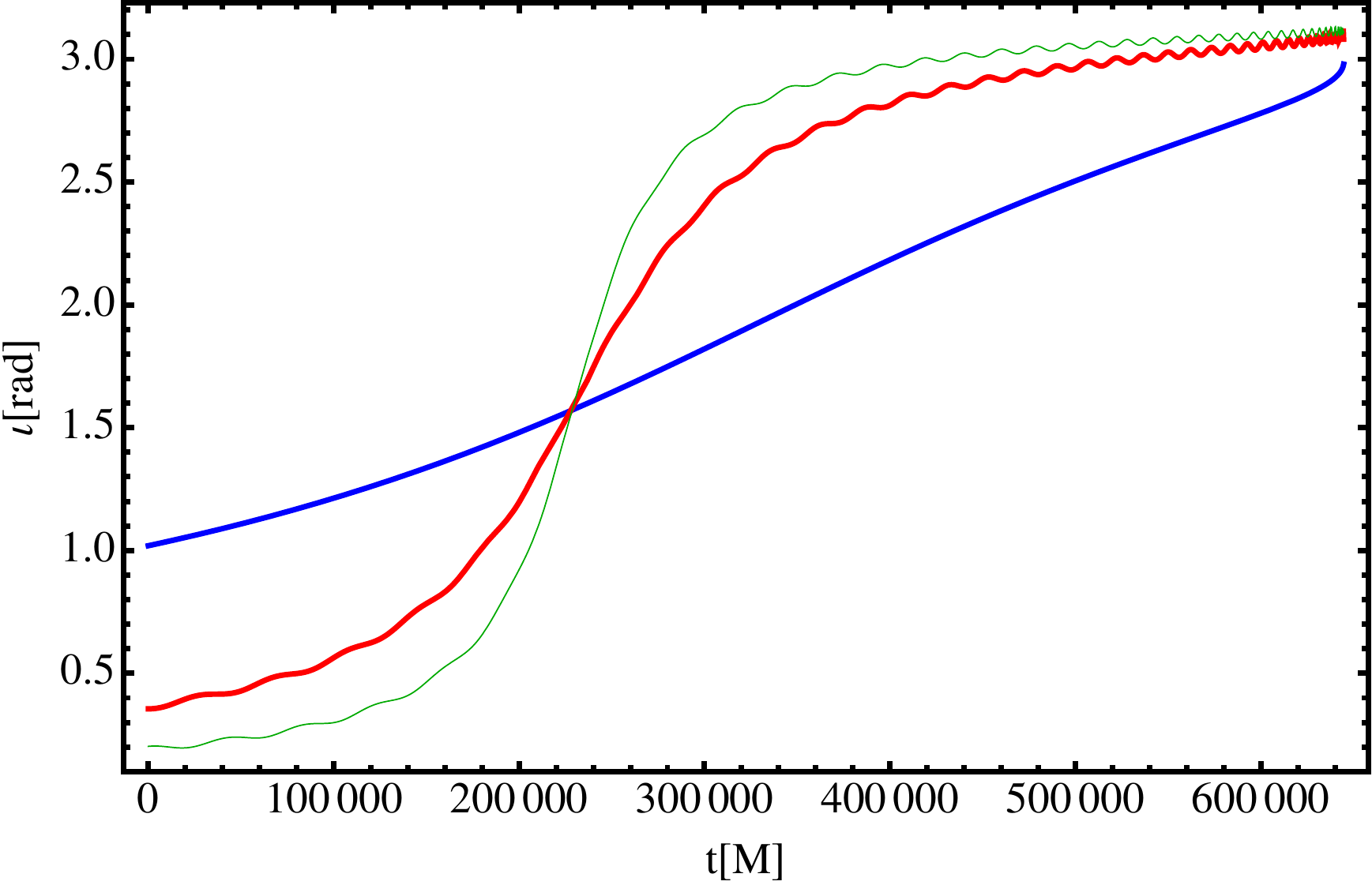}
\caption[Comparison of the precession angles $\iota$ and $\alpha$ for a case of transitional precession]{The left panel shows the PN evolution of the precession angle $\alpha$ for the transitional precession case described in the text (red) as well as $\alpha(t)$ for the corresponding reduced-parameter template (blue). The right panel compares the two precession cone opening angles. It is clear from those graphs that the mapping does not faithfully reproduce transitional precession. The green curves show the angles for a reduced-parameter system, where the precession is associated with the smaller black hole $m_1$, which appear to be closer to the angles in the generic system (red).
}
\label{fig:anglesTP}
\end{center}
\end{figure*}

The other class of possibly problematic cases that we identified in
Sec.~\ref{sec:chip} are those where precession is dominated by the small black
hole. We find that a total of 1699 of such configurations (17\%) for mass ratio
$q=3$. However, only $4.7\%$ among these matches are below the threshold
$\mathscr M = 0.965$.  Further, we find that these sub-threshold matches are
predominantly clustered around values for $\chi_{2\perp}\leq 0.08$. We conclude
that $\chi_p$ faithfully represents binaries that are precession-dominated by
the smaller black hole --- only systems with very little precession are not
faithfully approximated for certain binary inclination and signal polarisation
angles.

We expect the mapping onto the reduced-parameter waveforms to be yet even more
faithful for higher mass ratios such as $q=10$. On the other hand, we now expect
transitional precession to occur more often within the sensitivity band of
aLIGO. In order to identify the occurrence of transitional precession, we
following Ref.~\cite{Apostolatos:1994mx} and define an initial angle 
$\measuredangle{(\hat{L},\hat{S})}\geq 164^\circ$ to indicate transitional
precession. We find that $1.8\%$ of the sample configurations undergo the
transitional phase either completely or partially in band.

We again illustrate the results through the \CDF of matches in the purple
dot-dashed curve in Fig.~\ref{fig:q1CDF}. As expected, the tail is much flatter
than for the low mass ratio end with a fraction of only $0.3\%$ of all matches
below threshold. More than $97\%$ of cases show a match of $0.99$ or better. The
minimum match obtained is $\mathscr{M}_\mathrm{min}=0.484$ for the configuration
$\vec \chi_1 =-(0.56, 0.48, 0.06)$ and
$\vec \chi_2 =\{0.01, 0.02, -0.60\}$, which undergoes the full
transitional phase in band. The final angle between $\hat{J}$ and $(0,0,1)$ is
$146.6^\circ$. We illustrate the details of this particular case in the next
section.

\subsubsection{Special case: transitional precession}
Our random distribution of $q=10$ configurations includes some instances of in-band transitional precession.
As expected, these cases give, for certain orientations and polarisations, matches significantly below threshold, 
some as low as $\sim 0.4$.

Transitional precession occurs when the total spin $\vec{S}$ and the orbital
angular momentum $\vec{L}$ have similar magnitude but are directed nearly
opposite, such that the magnitude of the total angular momentum $J$ is small.
This will only occur within the frequency band of ground-based \GW detectors for
a narrow range of physical parameters. For a small set of configurations, the
binary starts in a simply precessing phase, then undergoes a transitional phase,
and, if it has not yet merged, returns to a state of simple precession. 

Fig.~\ref{fig:anglesTP} shows in red the evolution of the precession angles
$(\iota(t),\alpha(t))$ for the transitional configuration described previously.
The true physical system has initial spins $\vec \chi_1 = -(0.56, 0.48, 0.06)$
and $\vec \chi_2 = (0.01, 0.02, -0.60)$, while the corresponding
reduced-parameter configuration has initial spins $\vec{\chi}_1= -(0,0,0.06)$
and $\vec{\chi}_2=(0.06,0, -0.60)$. The comparison of the two precession angles
$\alpha$ and $\iota$ from the transitional configuration with its corresponding
model configuration reveals a strong disagreement. This can be explained as
follows: for transitional precession to also occur in the reduced-parameter
configuration, it is crucial that the parallel component of the total spin is
close to $S_{||}$ in the generic configuration. Since we fix the parallel spin
components in the mapping, the fulfilment of this condition is guaranteed. At
the same time, however, $S_{\perp}$ must also be similar to the full-parameter
system. If it is too large, the transitional phase occurs at later times; if it
is too small, the transition is shifted to earlier times. By construction,
$\chi_p$ corresponds to an average in-plane spin, which does not necessarily
correspond to $S_\perp$ of the generic system. We conclude that the faithful
representation of transitional precession is highly sensitive to the initial
value of $S_\perp$, but note that a different value of $\chi_p$ is in principle
capable of capturing transitional precession. In the green curves in
Fig.~\ref{fig:anglesTP} we illustrate this by placing the precession spin on the
smaller black hole, but similar results could also be achieved by optimising
over $\chi_p$ in our standard construction.

\subsection{On the goodness of $\chi_p$}

\begin{figure*}
\begin{center}
\includegraphics[width=80mm]{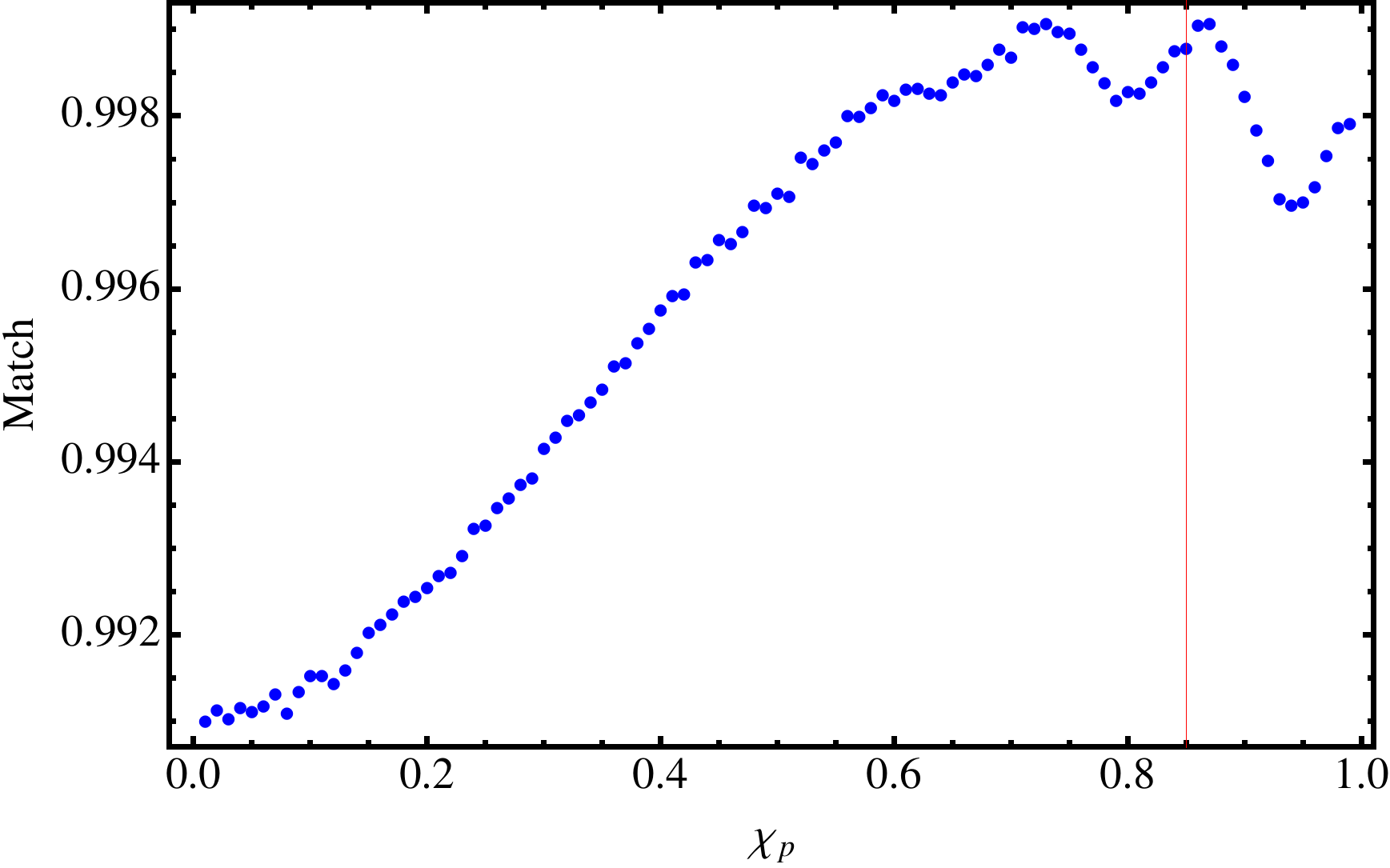} 
\quad
\includegraphics[width=80mm]{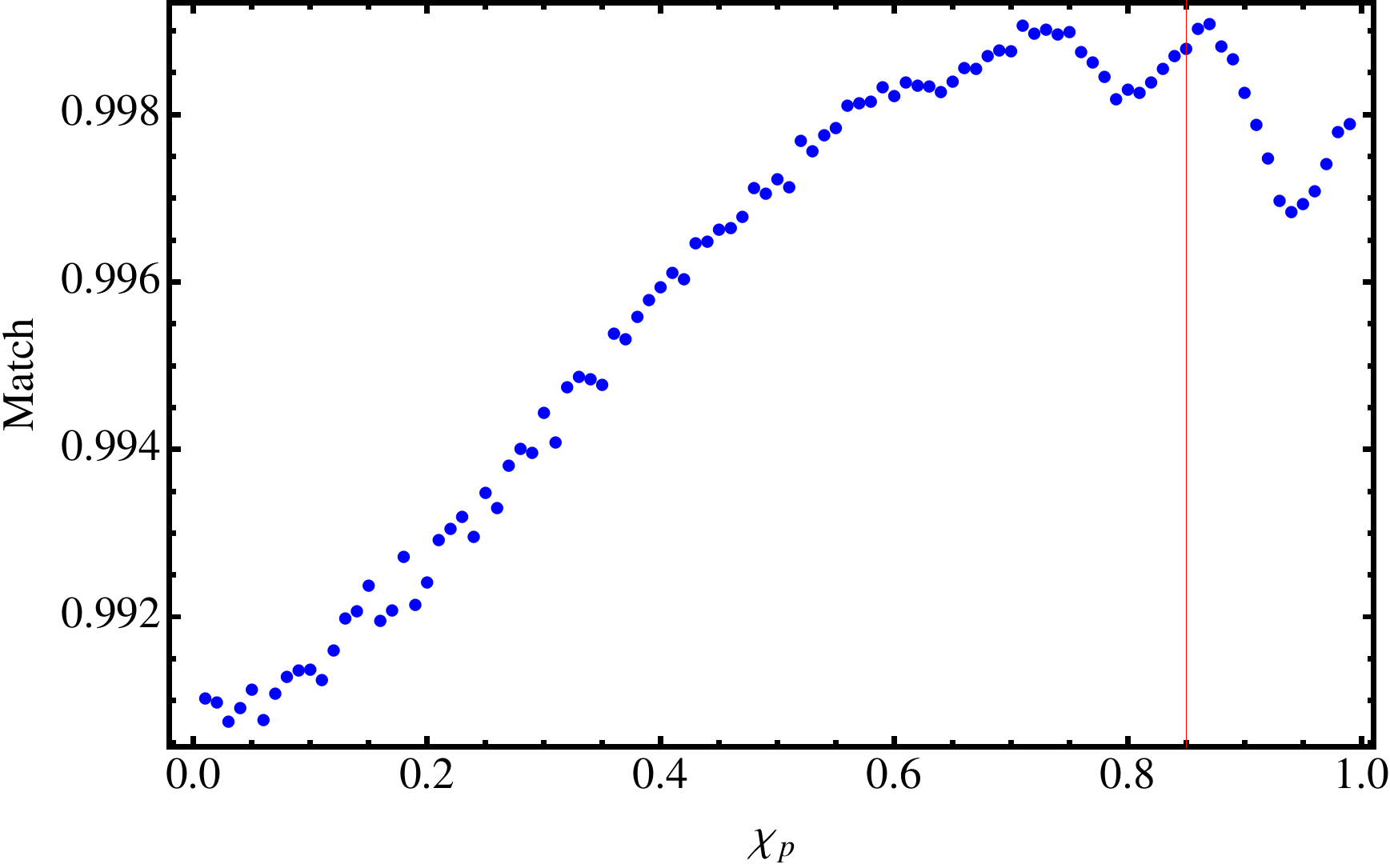}
\includegraphics[width=80mm]{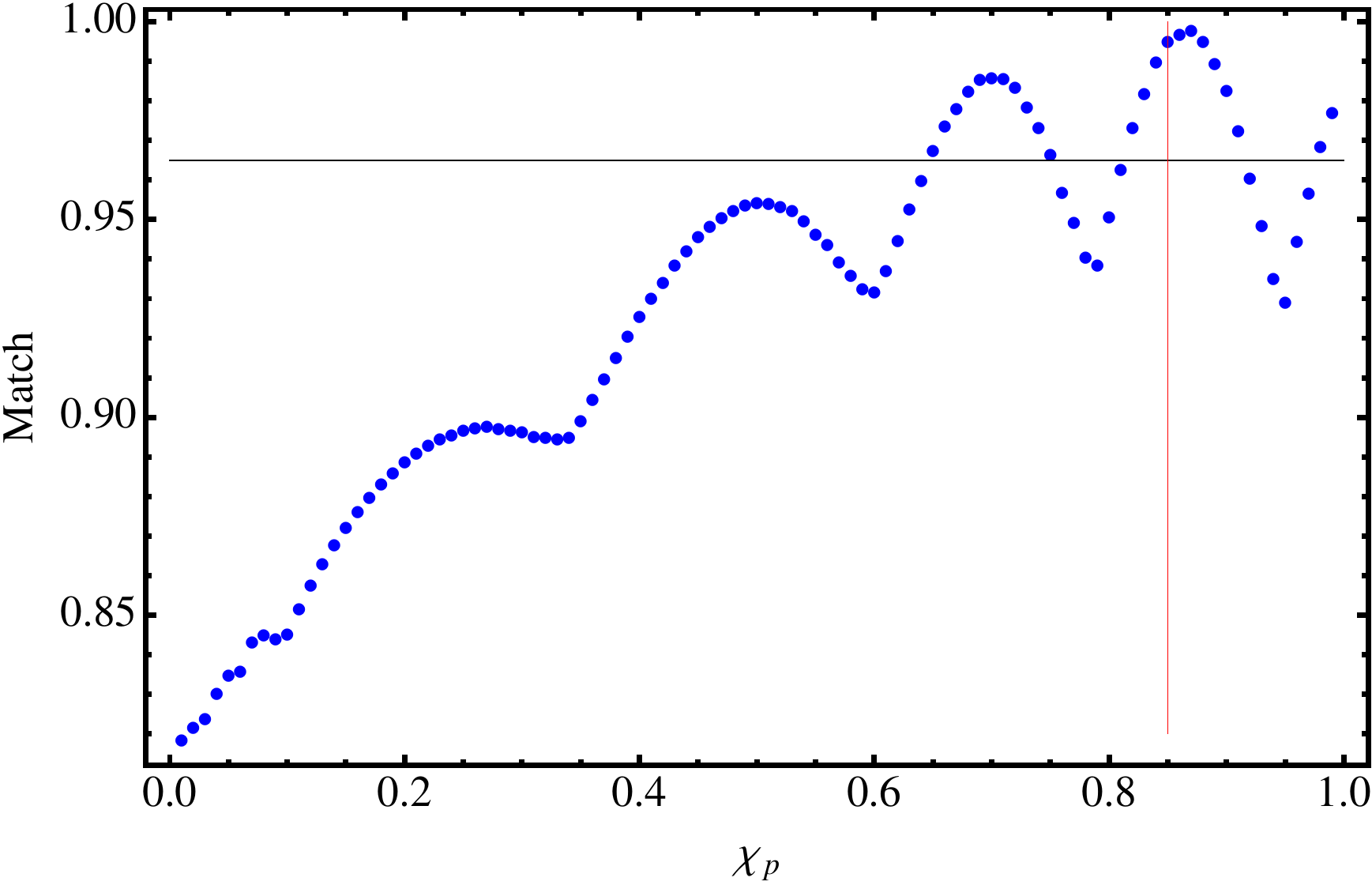}
\quad
\includegraphics[width=80mm]{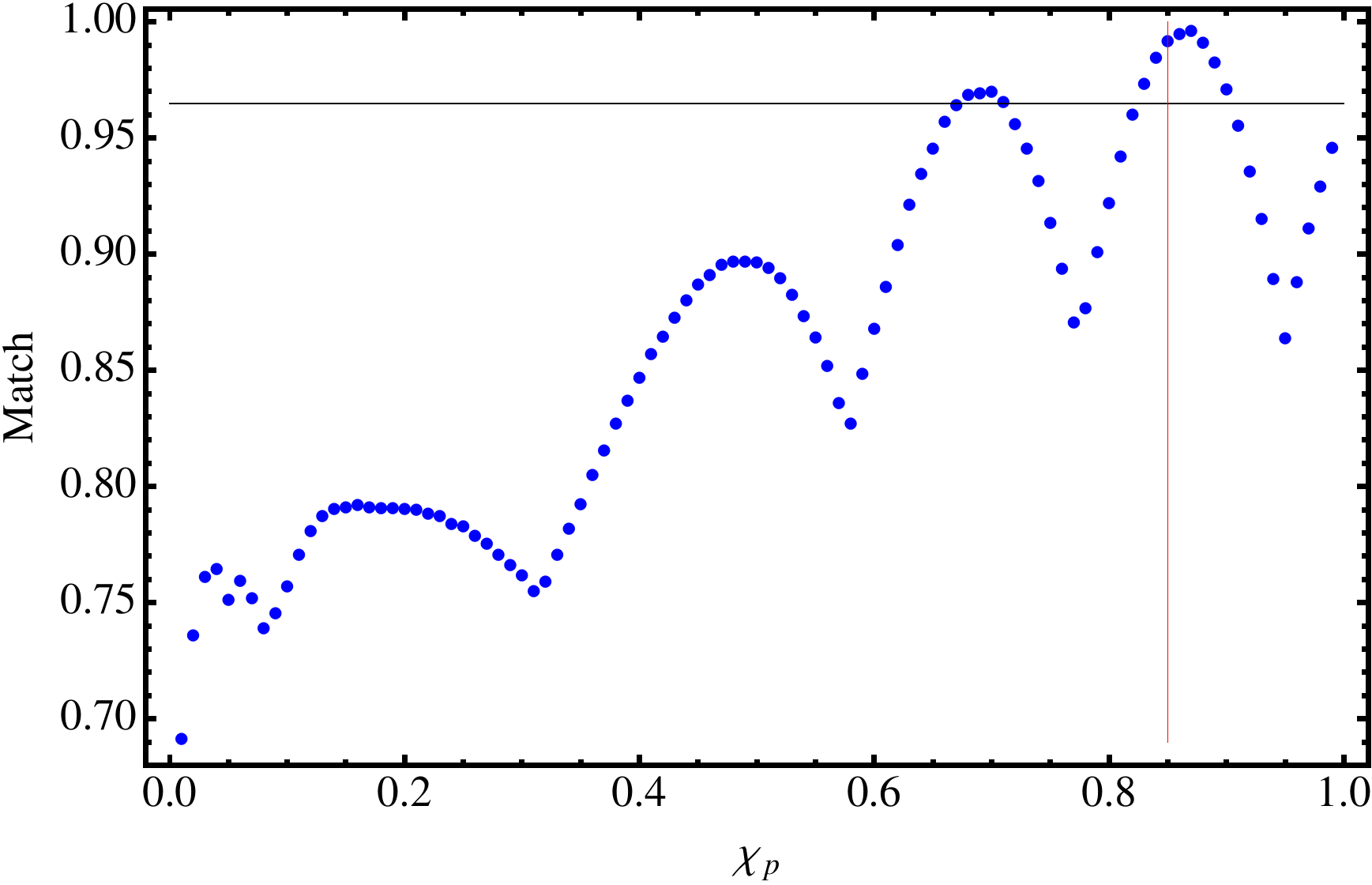}
\caption{The four panels show the matches for the case depicted in
Fig.~\ref{fig:AngleComp} with a series of reduced-parameter configurations with
varying $\chi_p$ for four different pairs of binary orientation and signal
polarisation $(\theta,\psi_S)$; these are from the top left to the bottom right:
$(0,\pi/8)$, $(0,\pi/4)$, $(\pi/4,\pi/8)$ and $(\pi/2,\pi/8)$.The red vertical
line indicates the theoretical $\chi_p$-value; the black horizontal line in the
lower two panels indicates the threshold of $\mathscr{M}=0.965$. We find a
strong dependence of the match on the value of $\chi_p$ for growing
inclinations, where waveform modulations become more pronounced. Moreover, the
theoretical $\chi_p$-value is very close to the value yielding the maximal
match.}
\label{fig:MatchChiP}
\end{center}
\end{figure*}

The results obtained so far suggest that the single spin parameter $\chi_p$
faithfully represents the precession in a given generic double-spin system. What
we have not yet investigated, however, is the goodness of this parameter, i.e.,
whether the particular definition of $\chi_p$ that we have chosen is ideal. We
can investigate this by determining the magnitude of the large black hole's
in-plane spin that yields the best agreement with the signal waveform, and
compare that with our estimate of $\chi_p$. To do so, we determine the match of
a single generic case with a series of reduced-parameter configurations, where
we vary the value of $\chi_p$. Previously, we have seen that the match strongly
depends on the inclination $\theta$ of the binary as well as the polarisation
angle $\psi_S$ of the signal. We therefore repeat the analysis for several
values
of $\theta$ and $\psi_S$. The results are illustrated in
Fig.~\ref{fig:MatchChiP}
for the same configuration as depicted in Fig.~\ref{fig:AngleComp},
$\{q=3,\vec{\chi}_1=(0.4,-0.2,0.3),\vec{\chi}_2=(0.75,0.4,-0.1)\}$.

For an optimally oriented binary (i.e., $\theta=0$) the match depends only
weakly on the explicit value of $\chi_p$. This is consistent with our
expectation that a large fraction of optimally-oriented precessing binaries is
well represented by aligned-spin binaries~%
\cite{Ajith:2009bn,Ajith:2011ec,Harry:2013tca,Ajith:2012mn,Brown:2012gs}. For
larger inclinations $\theta$, however, the match becomes more strongly dependent
on $\chi_p$. We find that the best match is indeed obtained for a $\chi_p$-value
close to the theoretically predicted one, indicating that $\chi_p$ does provide
a meaningful parameterisation of the precession and allows for a faithful
representation of a generic system in particular for large inclinations. This
needs to be investigated in more detail for a larger number of precessing
configurations, and for a full optimisation over all physical parameters, which
we defer to future work. 

Note the different scales in the upper and lower panels of
Fig.~\ref{fig:MatchChiP}. These suggest that our ability to measure
precession effects will depend strongly on the orientation of the binary. For
binaries with $\theta \approx 0$, it will be more difficult to distinguish that
a binary is precessing, than for one with larger orientations. Our ability to
measure not just whether the binary is precessing, but the value of $\chi_p$,
will of course also depend on all correlations of $\chi_p$ with other parameters
(which we keep fixed here) and on the errors in the waveform model, but this
requires a more in-depth study. 

It follows from the discussion after Eq.~(\ref{eq:Sp}) that, if we were able to 
accurately measure $\chi_p$, then for most configurations this would translate
into an accurate measurement of the in-plane spin magnitude of the larger black hole. 
Conversely, the in-plane spin of the smaller black hole would be poorly constrained. This is consistent
with the results given in Ref.~\cite{Vitale:2014mka}, where the spin of the larger black hole is 
in some cases measured to within 10\%.

\section{Comparison with other reduced-parameter families}
\label{sec:families}

In this work we have considered one choice of parameter-reduced waveform family,
i.e., we have replaced the black-hole spin components that lie in the orbital
plane (at some arbitrarily chosen time), with a binary with the same physical
parameters, except that now only the larger black hole has any in-plane spin,
and its magnitude is $\chi_p$, the effective precession spin parameter we
defined in Sec.~\ref{sec:chip}. The purpose of this study has been to determine
whether the dominant precession effects can be captured by a single ``precession
parameter", and our results suggest that in most cases it can. 

We can also infer from these results that our reduced-parameter waveform family
may be a good candidate for use in template banks in a search for precessing
binaries. We defer a detailed study of the efficacy of the $\chi_p$ waveforms in
searches to future work; in particular, such a study would require calculations
of fully optimised matches (fitting factors). However, it is natural to ask how
the $\chi_p$ family compares to reduced-parameter families that have been
suggested in previous work, or how those parameter reductions might be combined
with our $\chi_p$ approximation. 

We consider three families: the single-spin ``Physical Template
Family"~\cite{Buonanno:2004yd}, and two waveform families that also use the
``effective spin'' approximation to reduce the two spin components parallel to
the orbital angular momentum to a single parameter, $\chi_\mathrm{eff}$.

\subsection{Comparison with the \emph{Physical Template Family}}

Buonanno et al.~\cite{Buonanno:2004yd} suggested in 2004 a single-spin
precessing waveform family that is effectual in detecting generic double-spin
precessing binaries. Their quasi-physical template family (PTF) exhibited very
high fitting factors across a wide range of configurations. We do not calculate
fitting factors here, and therefore cannot make a direct comparison with PTF,
but by comparing our partially optimised matches with PTF will give us an
indication of how they may compare in terms of parameter estimation.

Let us first point out the differences between the two waveform families. Based
on the approximate decoupling between the inspiral and precession dynamics, we
suggest that the inspiral is well described by the two parallel spin components,
whereas the precession can be encapsulated in a single complementary spin
parameter. This yields a double-spin system with three spin parameters as given
in Eq.~(\ref{eq:map}). PTF, on the other hand, assigns the total spin $\vec S$
of the double-spin configuration to the larger black hole, resulting in a pure
single-spin system, again with three spin parameters, obtained by the following
map:
\begin{align}
\label{eq:mapPTF}
   \vec{\chi}_1 & \mapsto (0,0,0),   \\
    \vec{\chi}_2& \mapsto \frac{\vec{\chi}_1m_1^2+\vec{\chi}_2 m_2^2}{m_2^2} . 
\end{align}

This mapping can be compared with the reduced-parameter mapping we use in
Eq.~(\ref{eq:map}). We expect that our mapping will allow us to correctly
capture the inspiral rate (through the two parallel spin components), while
$\chi_p$ will drive the appropriate precession. In contrast, the PTF mapping
provides only one parallel spin component, and we therefore expect that it will
not capture the inspiral rate so accurately. 

\begin{figure}
\includegraphics[width=80mm]{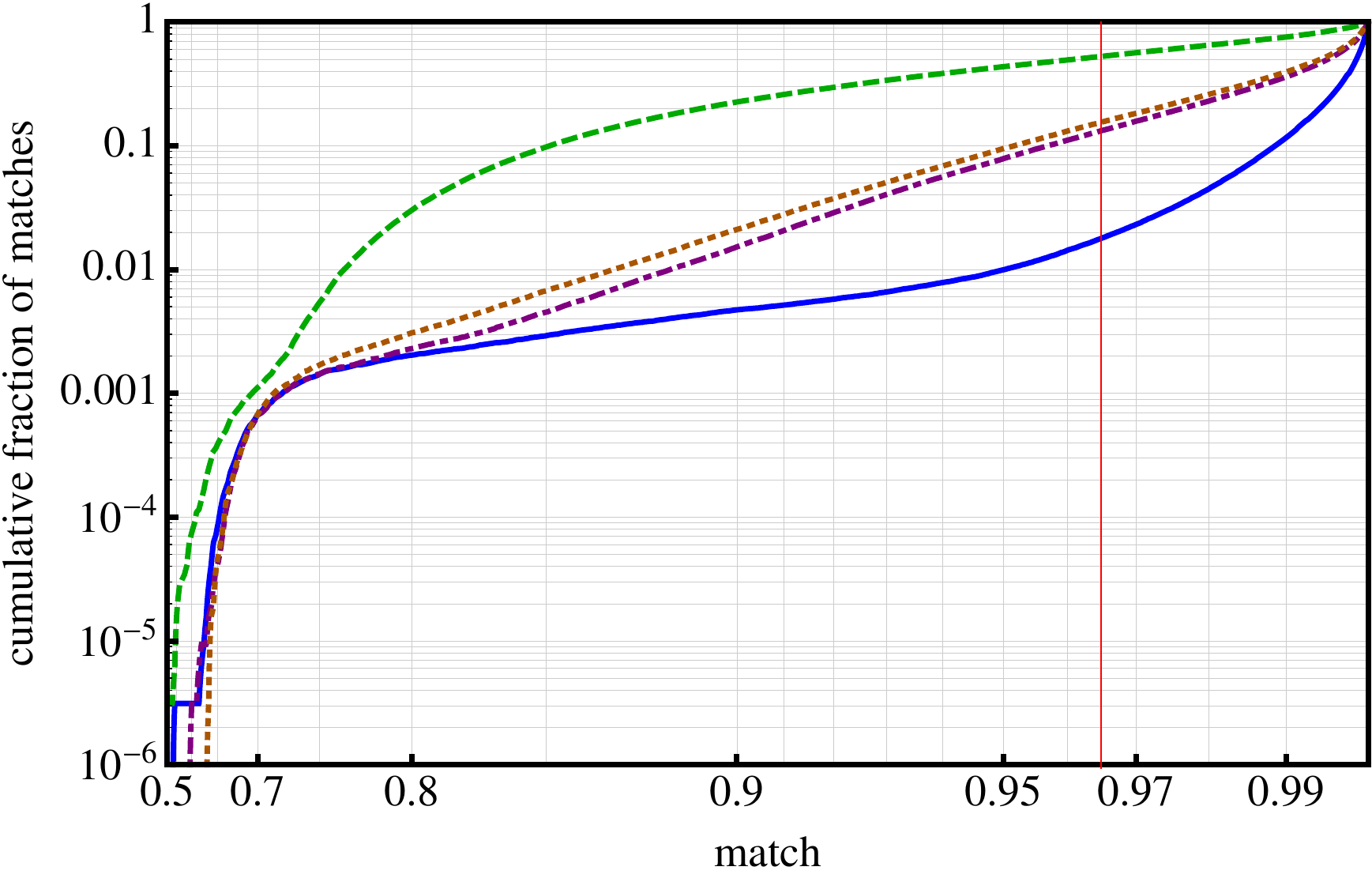}
\caption{Comparison of the $\chi_p$ model with alternative parameter reductions.
Our $q=3$ results from Fig.~\ref{fig:q1CDF} are repeated in the blue curve.
Also shown are the PTF parameterisation (green, dashed), the single parallel
spin $\chi_\mathrm{eff}$ applied to the larger BH only (purple, dot-dashed) and
$\chi_\mathrm{eff}$ as parallel spin on both BHs (orange, dotted). See text for
more details.}
\label{fig:PTF}
\end{figure}

We now test that conjecture by calculating the match of both approximations for
one comparable mass ratio $q=3$ using the same sample of generic spin
configurations as in Sec.~\ref{sec:stats}. We apply our proposed mapping to each
configuration, as well as the PTF  mapping, and compute the matches against the
double-spin target signal, respectively. Fig.~\ref{fig:PTF} shows the cumulative
distribution function for both mappings. We find that the mapping suggested by
PTF results in $53\%$ of all matches smaller than $0.965$, compared to only
$\sim2\%$ for the mapping given in Eq.~(\ref{eq:map}). We therefore conclude
that the assignment of the total spin to the larger black hole does not yield a
particularly faithful representation of the generic double-spin system, whereas
the split into the parallel spin components $\chi_{i||}$ and $\chi_p$ yields
matches above threshold for $\sim 98\%$ of all configurations.

Another way to interpret this result is that one of the three spin components in
the PTF mapping is the orientation of the larger black hole's spin in the
orbital plane. This orientation is approximately degenerate with the binary's
orientation angle $\varphi_S$, and any variation in this angle has only a small
effect after optimising over the corresponding model angle $\varphi_M$. This
leaves the PTF model with only two other spin parameters with which to capture
the waveform, while our $\chi_p$ model has three. 

We shall investigate in the next section whether a parameter reduction
from three to two spin parameters completely accounts for the loss in 
accuracy observable in Fig.~\ref{fig:PTF} for the PTF model.

\subsection{$\mathbf{\chi_\mathrm{eff}}$-parameterisation of the inspiral rate}
\label{sec:chieffpara}
In the analysis presented so far, we have kept the parallel spin components in
the reduced-parameter system the same as in the full-parameter system. We now
investigate an additional parameter reduction, where we now replace the two
parallel spin components with the effective inspiral spin, $\chi_\mathrm{eff}$,
as used in several phenomenological \IMR
models~\cite{Ajith:2009bn,Santamaria:2010yb}, including a precessing-binary
model~\cite{Hannam:2013oca}; its efficacy was studied in more detail in
Ref.~\cite{Purrer:2013ojf}. Since we use \PN inspiral waveforms, we use the
extended version of the effective inspiral spin as defined
by~\cite{Ajith:2011ec},
\begin{equation}
\chi_{\rm eff} = \frac{1}{2}\left( 1-
\frac{76\eta}{113}\right)\left(\chi_{1||}+\chi_{2||}\right)+\frac{1}{2}\frac{
\delta M}{M}\left( \chi_{1||}-\chi_{2||}\right), \label{eq:chieff}
\end{equation}
where $\delta M=m_1-m_2$ and $\eta=m_1m_2/M^2$ the symmetric mass ratio.

We consider two parameter reductions with $\chi_{\rm eff}$. One is to place all
of the spin on the larger black hole. If we once again define our configurations
with respect to $\hat{L} \equiv \hat{z}$, this reduction is then given by:
\begin{equation} 
\label{eq:effMap} \vec{\chi}_1\mapsto (0,0,0), \quad \vec{\chi}_2 \mapsto \left(\chi_p,0,\frac{113 \chi_\mathrm{eff}}{(113-76 \eta)}\right). 
\end{equation} 
This is the choice that is made in the construction of the PhenomP
precessing-binary model~\cite{Hannam:2013oca}. The second option is to choose
\begin{equation}
\label{ }
\chi_{i||}=2 \chi_\mathrm{eff}\left(1-\frac{76 \eta}{113}-\frac{\delta M}{M}\right)^{-1}
\end{equation}
as the parallel spin component on each black hole. 

We investigate the faithfulness of these two parameterisations using the same
$q=3$ configurations as in Sec.~\ref{sec:stats}. Fig.~\ref{fig:PTF} illustrates
the results in the form of the cumulative fraction of matches as a function of
the match. We see that both $\chi_{\rm eff}$ models show an improved 
performance compared to PTF, but a worse performance than the original
three-spin-parameter model. We now find $\sim 14\%$ and $\sim
16\%$ of all matches below the threshold, respectively.
With the reduction of the parallel spin components,
both the accuracy in modelling the secular phasing is slightly decreased, and the
precession dynamics is also affected, as it is governed by the total spin (in particular the initial 
precession cone opening angle) rather than
the effective spin combination. The sum of both effects leaves us with $\sim 15\%$ of matches 
below threshold.

\section{Discussion}
\label{sec:conclusion}

In the analysis presented here, we have explored the possibility of
parameterising the precession in generic double-spin black-hole binaries with
only one precession spin parameter. The inspiral and precession
dynamics approximately decouple, and the precession is determined predominantly
by the spin components that lie in the orbital plane. The leading-order
precession effects in \PN theory indicate that these in-plane spin components
rotate in the orbital plane at different rates and that their magnitudes show
only small variations. 
This motivates a spin parameter,
$\chi_p$, which is defined as a simple mean of the in-plane spins. 
   
We have tested the effectiveness of this parameterisation by constructing a
reduced-parameter family of binary waveforms, where we replace the in-plane spin
components by assigning $\chi_p$ as the in-plane spin of the larger black hole (see Eq.~(\ref{eq:map})).
We have quantified the accuracy of these reduced-parameter waveforms for
the extreme case of one or two maximally spinning black holes, with respect to variations
of the relative in-plane spin orientations (Sec.~\ref{sec:orientation}), the in-plane spin 
magnitude (Sec.~\ref{sec:magnitude}), and the magnitude of the spin components parallel 
to the orbital angular momentum that affect the inspiral rate
(Sec.~\ref{sec:parallel}).
In most cases the reduced-parameter waveforms agree well with the 
full-parameter signals, with the worst agreement occurring when the line-of-sight
between the source and the detector is perpendicular to the binary's total
angular momentum. We have identified this disagreement to mainly originate
from partially neglecting spin-spin 
interactions in our model.

We compared our waveform family against a random sample of 10,000
configurations at mass ratios $q = 1,3, 10$. The agreement of our
reduced-parameter model with each full-parameter configuration is shown in
Fig.~\ref{fig:q1CDF} which indicates good agreement for a large fraction of
configurations. This is even true in the equal-mass case, where the in-plane
spins rotate at the same rate, and therefore averaging over the spin orientation
becomes invalid and $\chi_p$ no longer approximates the true average precession
rate. However, the error in this approximation appears to be no greater than the
error in neglecting spin-spin effects. 
 
The efficacy of the precession parameter has implications for \GW measurements.
If the dominant precession effects can be captured with only one spin parameter, then this
indicates that it will be difficult to distinguish the individual spin vectors in a GW observation.
We already know that if the binary's total angular
momentum is oriented towards the detector, then the precession will have only a
minimal effect on the waveform, and so the precession will be difficult to
detect. But even in binaries where the total angular momentum has a large
inclination angle with respect to the detector's line-of-sight, and precession
effects are strong, it may be difficult to identify both of the individual in-plane spin
magnitudes. However, as discussed following Eq.~(\ref{eq:Sp}), for many configurations, 
if we can accurately measure the parallel spin components and $\chi_p$, then we will be
 able to accurately measure the spin magnitude of the larger black hole.

The results of this paper add to our overall understanding of
the dominant parameters
that \emph{will} be measurable in \GW observations of binary coalescences. In
aligned-spin 
binaries we can most accurately measure a combination of the component masses 
(the chirp mass)~\cite{Arun:2004hn}. At the next level of accuracy, we can measure a 
combination of the binary's mass ratio and a combination of the parallel components of 
the black-hole masses --- but not, at moderate signal-to-noise ratios, the individual black-hole 
spins~\cite{Cutler:1994ys,Poisson:1995ef,Baird:2012cu,Hannam:2013uu,Ohme:2013nsa,Purrer:2013ojf}. 
To this picture
we add the precession parameter $\chi_p$, which tells us that for the in-plane spin components,
it is only one of them that we will most likely be able to measure. 

How well we can measure each of these parameters, and what configurations 
allow us to sufficiently break the 
degeneracies in order to estimate both of the individual black-hole
spins, will depend not only on the signal-to-noise ratio of the signal, but also
on the binary configuration and its
relative orientation to the detector. This is a topic that deserves further study in the future, 
building on the work already done in Ref.~\cite{Vitale:2014mka}.

Based on our results, we suggest that a waveform model with three spin parameters, which
uses the same parameter reduction as in Sec.~\ref{sec:chip}, may be more
effective in \GW detection and parameter estimation than alternative
parameterisations, for example the PTF reduction suggested in
Ref.~\cite{Buonanno:2004yd}. As we discuss in Sec.~\ref{sec:families}, this is
because the parameter reduction we propose accurately models separately 
the inspiral rate
(using the two parallel spin components), as well as the dominant precession
dynamics (using $\chi_p$). One could also consider a parameter reduction based on
only two spin components, one for the effective parallel spin, and another for the
effective precession spin. We find that these models perform well, with a
significant improvement over the PTF mapping to single-spin systems.  

The present study has been limited to inspiral waveforms only, and has not
included match comparisons that are optimised over the source parameters; we have
also neglected the effect of higher PN order spin terms, which may weaken the 
$\chi_p$ degeneracy that we have identified. The
purpose here was to demonstrate the utility of a single precession parameter in
capturing the average precession exhibited by a generic binary system.
More detailed studies are required to determine the value of this parameter
reduction in \GW searches and in parameter estimation, and in waveforms that
include merger and ringdown. Depending on the extent to which this partial
degeneracy holds throughout the entire \IMR waveform, it may be possible to
accurately model generic binaries with NR simulations that cover a reduced
parameter space, thus making far more tractable the problem of constructing
generic \IMR models for use in \GW astronomy with Advanced detectors. However,
the identification of the dominant physical parameters in the inspiral is valuable in
itself in simplifying the construction of precessing waveform models and in particular for
producing a sufficient analytic description of the rotation that describes the
evolution 
of the orbital plane; for example, 
the frequency-domain precessing IMR model proposed in Ref.~\cite{Hannam:2013oca} 
was motivated in part by a preliminary version of the results presented here.

\section*{Acknowledgments}
We thank G. Faye for providing us with a Mathematica notebook containing all 
waveform-mode expressions used in this work.
We also thank A. Boh\'e, S. Fairhurst, S. Husa, G. Pratten, M. P\"{u}rrer and B. Sathyaprakash
for valuable 
discussions and for comments on the manuscript.
P. Schmidt is a recipient of a DOC-fFORTE-fellowship of the Austrian Academy of Sciences
and was also partially supported by the STFC.
M. Hannam was supported Science and Technology Facilities Council grant ST/H008438/1, and
both M. Hannam and F. Ohme by ST/I001085/1.
%
\appendix
\allowdisplaybreaks
\section{PN waveform generation}
\label{sec:wfgen}
For efficiency reasons, the PN waveforms used in the analysis presented here are generated by integrating the 2.5PN orbit-averaged precession equations under the assumption of quasi-spherical inspiral for $\vec{L}$ and $\vec{S}_i$ as given in~\cite{Kidder:1995zr}:
\begin{eqnarray}
\dot{\vec{L}}&=&\frac{1}{r^3}\left[ \left ( 2 +\frac{3q}{2}\right)\vec{S}_1+
\left( 2+\frac{3}{2q}\right) \vec{S}_2\right] \times
\vec{L} \nonumber\\ \label{eq:Ldot} 
&& -\frac{3}{2r^3}\left[ (\vec{S}_2\cdot
\hat{L})\vec{S}_1+(\vec{S}_1\cdot \hat{L})\vec{S}_2\right] \times
\hat{L} \\
&&-\frac{32\mu^2}{5r}\left( \frac{m}{r}\right)^{5/2}\hat{L}, \nonumber \\
\label{eq:S1dot}
\dot{\vec{S}}_1 &=& \frac{1}{r^3}\left[  \left ( 2 +\frac{3q}{2}\right) \hat{L}
+ \frac{1}{2}\vec{S}_2 -\frac{3}{2} (\vec{S}_2 \cdot \hat{L})\hat{L} \right]
\times \vec{S}_1, \\
\label{eq:S2dot}
\dot{\vec{S}}_2 &=& \frac{1}{r^3}\left[  \left ( 2 +\frac{3}{2q}\right) \hat{L}
+ \frac{1}{2}\vec{S}_1 -\frac{3}{2} (\vec{S}_1 \cdot \hat{L})\hat{L} \right]
\times \vec{S}_2.
\end{eqnarray}

The evolution equation for the precession angle $\alpha(t)$ is determined
by differentiating Eq.~(\ref{eq:alpha}) with respect to time,
\begin{align}
\label{eq:alphadot}
 \dot{\alpha}(t) =& \frac{L_x\dot{L}_y-L_y\dot{L}_x}{L_x^2+L_y^2+\epsilon},
\end{align}
where $\epsilon=10^{-4}$ to ensure that the expression does not diverge in the
numerical integration. The opening angle $\iota(t)$ is obtained from
Eq.~(\ref{eq:iota}).

Further, we integrate the evolution equation for the
orbital separation, $r$, and construct the orbital frequency, $\omega_\mathrm{orb}$, from it:
\begin{widetext}
\begin{eqnarray}
\dot{r}(t) &=&-\frac{64\eta}{5}\left (
\frac{m}{r}\right )^3\left[ 1-\frac{1}{336} (1751+588\eta) - \bigg\{ \frac{7}{12} \sum_{i=1,2} \left[ \chi_i (\hat{L} \cdot \hat{S}_i) \left ( 19\frac{m_i^2}{m^2}+15 \eta\right)\right] -4\pi \bigg\} \left (\frac{m}{r} \right )^{3/2} \right. \nonumber \\ 
&& \left. -\frac{5}{48}\eta\chi_1\chi_2\left[ 59(\hat{S}_1\cdot
\hat{S}_2)-173(\hat{L}\cdot \hat{S}_1)(\hat{L}\cdot \hat{S}_2)\right]\left (
\frac{m}{r}\right )^2
\right], \label{eq:rdot} \\
\omega_\mathrm{orb}^2 &=& \left(\frac{m}{r^3}\right) \bigg \{
1-(3-\eta)\left(\frac{m}{r}\right)-\sum_{i=1}^{2} \left[ \chi_i (\hat{L}\cdot
\hat{S}_i) \left( 2\frac{m_i^2}{m^2}+3\eta \right)\right]
\left(\frac{m}{r}\right)^{3/2}+\left[ \left(6+\frac{41}{4}\eta+\eta^2 \right)
\right. \nonumber \\
& & \left.-\frac{3}{2}\eta\chi_1\chi_2\left[ (\hat{S}_1\cdot
\hat{S}_2)-3(\hat{L}\cdot \hat{S}_1)(\hat{L}\cdot \hat{S}_2)\right] \right]
\left( \frac{m}{r}\right)^2 \bigg \}. \label{eq:omega}
\end{eqnarray}
\end{widetext}

We then integrate the equation for the total phase, Eq.~(\ref{eq:totalphase}).
The evolution is performed in the $J_0$-aligned frame and is terminated
when a final separation of $r=6M$ (corresponding to the last stable
circular orbit in the Schwarzschild spacetime) is reached. As initial
conditions we choose the spin components defined with respect to $\hat{L}_0
\equiv (0,0,1)$, the initial separation $r_0=40M$, the initial orbital phase
$\Phi_0=0$ and the initial azimuth of $\hat{L}$ in the $J_0$-aligned frame. We
also have to set the initial magnitude of the orbital angular momentum, which we
choose to be the Newtonian value, $L_0 \equiv L_N = m_1 m_2 \sqrt{r_0/M}$. The
transformation into the $J_0$-aligned frame is given by the following rotation
matrix:
\begin{equation}
\label{eq:rotmatrix}
\mathbf{R}=\mathbf{R}_z(\epsilon_0-\pi)\mathbf{R}_y(-\iota_0)\mathbf{R}_z(-\epsilon_0),
\end{equation}
where $\epsilon_0$ is the initial azimuth of the total angular momentum $J_0$.

Once we have solved for the dynamics of the binary, we use the mode expressions
$h_{\ell m}$ as given in~\cite{Arun:2008kb} to construct the precessing
waveforms. We only use the $(\ell=2)$-modes and truncate the amplitudes at
leading \PN order ($v^2$), yielding the following explicit mode expressions:
\begin{widetext}
\begin{align}
\label{eq:arunhlm}
&h_{22}=-\frac{A}{2}e^{-2i (\iota-\alpha-\Phi)}\left[ e^{4i\phi} \left (
-1+e^{i\iota}\right )^4+\left ( 1+e^{i\iota}\right)^4 \right], \\
&h_{21}=-iAe^{-i(\alpha+2\Phi+2\iota)}\left[ -e^{4i\Phi} \left(
1+e^{i\iota}\right)\left( -1+e^{i\iota}\right)^3-\left( 1+e^{i\iota}\right)^3
\left( -1+e^{i\iota}\right) \right], \\
&h_{20}=A\sqrt \frac{3}{2}e^{-2i(\iota+\Phi)}\left(
-1+e^{2i\iota}\right)^2\left( 1+e^{4i\Phi}\right), \\
&h_{2,-2}=-\frac{A}{2}e^{2i(\alpha+\Phi+\iota)}\left[ e^{-4i(\Phi+\pi)}\left(
-1+e^{-i\iota}\right)^4+\left( 1+e^{-i\iota}\right)^4\right], \\
&h_{2,-1}=iAe^{i(\alpha+2\iota+2\Phi+\pi)}\left[
-e^{-4i(\Phi+\pi)}\left(-1+e^{-i\iota} \right)^3 \left( 1+e^{-i\iota}\right)-
\left( -1+e^{-i\iota}\right) \left( 1+e^{-i\iota}\right)^3 \right],
\end{align}
\end{widetext}
where the amplitude factor is
\begin{equation}
A=\frac{M\eta}{D_L}v^2\sqrt\frac{\pi}{5}.
\end{equation}
In the above equations $D_L$ is the luminosity distance of the GW source which we set to $D_L=1$.

\section{Generic matches}
\label{sec:precmatch}
Following the notation introduced in Sec.~\ref{sec:gen_match}, we define the
real-valued detector response
\begin{eqnarray}
 \hr(t) &=& \cos(2\psi) \, h_+(t) + \sin(2\psi) \, h_\times(t) \\
     &=& \Re\left[ h(t) \; e^{ i2\psi} \right],
\label{eq:complex_polarisation_app}
\end{eqnarray}
where $\psi$ is the polarisation angle and
\begin{equation}
 h = h_+ - i \, h_\times
\end{equation}
is the complex \GW strain. As discussed in Sec.~\ref{sec:gen_match}, we remind
the reader that our definitions of $h_+$ and $h_\times$ include the 
orientation-dependent antenna pattern of the detector, except for the effect of
a relative rotation in the detector plain that is explicitly governed by $\psi$.

Our goal is to formulate the inner product between a signal and a model
response in terms of their complex \GW strains. We first express the
Fourier-domain detector response, $\fhr$, by
\begin{eqnarray}
 \hr(t) &=& \frac 12 \left[ h(t) \; e^{i2 \psi} + h^\ast(t) \; e^{-i2 \psi}
\right], \\
 \Rightarrow ~~  \fhr(f) &=& \frac 12 \left[ \tilde h(f)  \; e^{i2 \psi} +
\tilde h^\ast(-f) \; e^{-i2 \psi}
\right].
\end{eqnarray}
We then simply insert this expression into the inner product,
\begin{equation}
 \inner{\hr^S}{\hr^M} = 2 \int_{-\infty}^{\infty} \frac{\fhr^S (f) \; 
\fhr^{M\ast}(f)}{S_n(\vert f \vert)} d\!f, 
\end{equation}
and group the terms conveniently
\begin{widetext}
\begin{eqnarray}
 \inner{\hr^S}{\hr^M} &=& \frac 12 \int_{-\infty}^{\infty}  \left[ \tilde h^S(f)
\, \tilde h^{M \ast}(f) \, e^{i2(\psi_S - \psi_M)} + \tilde h^{S\ast}(-f)
\, \tilde h^{M}(-f) \, e^{-i2(\psi_S - \psi_M)} \right. \nonumber \\*
&& \qquad + 
\left. \tilde h^S(f)
\, \tilde h^{M}(-f) \, e^{i2(\psi_S + \psi_M)} + \tilde h^{S\ast}(-f)
\, \tilde h^{M\ast }(f) \, e^{-i2(\psi_S + \psi_M)} \right]
\frac{d\!f}{S_n(\vert f \vert)} . \label{eq:match_exp_app}
\end{eqnarray}
\end{widetext} %
Noting that 
\begin{equation}
 \int_{-\infty}^{\infty} x(f) \; d\!f = \int_{-\infty}^{\infty} x(-f) \; d\!f
\end{equation}
for any integrable function $x$, we identify the first and last two terms in
(\ref{eq:match_exp_app}) as complex conjugates of each other, respectively,
which leads to the final expression
\begin{eqnarray}
\inner{\hr^S}{\hr^M} = \mathrm{Re}
\int_{-\infty}^\infty \frac{\tilde h^S(f) \,
\tilde h^{M\ast}(f)}{S_n(\vert f \vert)} e^{2i(\psi_S-\psi_M)} d\!f \nonumber
\\*
 + \mathrm{Re} \int_{-\infty}^\infty \frac{\tilde h^S(f) \,
\tilde h^{M}(-f)}{S_n(\vert f \vert)}e^{2i(\psi_S+\psi_M)}d\!f.
\label{eq:fullmatch_app}
\end{eqnarray}
The first contribution in (\ref{eq:fullmatch_app}) closely resembles the
``standard'' formulation of the inner product, where the overall phase
difference is now identified as a difference of the polarisation angles. The
second term quantifies the asymmetry between positive and negative frequencies,
or, equivalently, the non-stationarity in the waveform strain.

In the following we are interested in the normalised match between the signal
and model, so we need to express norm of each waveform which, according to
 (\ref{eq:fullmatch_app}), reads
\begin{eqnarray}
 \| \hr \|^2 &=& \inner{\hr}{\hr} \\
  &=& \int_{-\infty}^\infty \frac{\vert \tilde h(f) \vert^2}{S_n(\vert f
\vert)} d\!f 
  + \mathrm{Re} \int_{-\infty}^\infty \frac{\tilde h(f) \,
\tilde h(-f)}{S_n(\vert f \vert)}d\!f e^{4i\psi}. \nonumber
\end{eqnarray}
Again, while the first term is similar to the standard norm of
non-precessing signals, there is a second (generally smaller) contribution that
quantifies the asymmetry and makes the norm polarisation dependent.

To find the optimal match over all polarisation angles of the model, $\psi_M$,
we rephrase the expressions above in terms of the real-valued quantities $N_1$,
$N_2$, $O$, $\sigma_N$ and $\sigma_O$,
\begin{eqnarray}
 N_1 &=& \int_{-\infty}^\infty \frac{\vert \tilde h^M(f) \vert^2}{S_n(\vert f
\vert)} d\!f , \nonumber \\
 N_2 e^{i \sigma_N} &=& \int_{-\infty}^\infty \frac{\tilde h^M(f) \,
\tilde h^M(-f)}{S_n(\vert f \vert)}d\!f , \label{eq:realconst_app} \\ \nonumber
O e^{i \sigma_O} &=& \int_{-\infty}^\infty 
\frac{\tilde h^{M \ast}(f)}{S_n(\vert f \vert)} \left[ \tilde
h^S(f)e^{2i\psi_S} + \tilde h^{S \ast} (-f) e^{-2i\psi_S} \right] d\!f.
\end{eqnarray}
This allow us to express the optimised match in the following way:
\begin{widetext}
\begin{eqnarray}
\max_{\psi_M}\inner{\frac{\hr^S}{\| \hr^S\|}}{\frac{\hr^M}{\| \hr^M\|}}
 &=& \max_{\psi_M} \frac{O}{\| \hr^S \|} \frac{\cos(2 \psi_M -
\sigma_O)}{\sqrt{N_1 +
N_2 \cos(4 \psi_M + \sigma_N)}} 
= \frac{O}{\| \hr^S \|} \sqrt{\frac{N_1 - N_2 \cos(\sigma_N + 2
\sigma_O)}{N_1^2
- N_2^2}} , \\
\psi_M^{\rm opt} &=& \frac12 \arctan \frac{N_1 \sin(\sigma_O) + N_2
\sin(\sigma_N
+
\sigma_O)}{N_1 \cos(\sigma_O) - N_2 \cos(\sigma_N +
\sigma_O)}.
\end{eqnarray}
\end{widetext}
Note that these expressions are understood as matches for constant signal
parameters (including $\psi_S$). However, re-computing the matches for a range
of signal polarisations $\psi_S$ is computationally cheap as only $O$ and
$\sigma_O$ have to be re-evaluated following (\ref{eq:realconst_app}).

The other parameter that we optimise over is a relative time shift between the
signal and the model, which enters the match (\ref{eq:fullmatch_app}) as a
complex modulation $e^{2\pi i f \Delta t}$. As usual, we efficiently calculate
the match for discretized time shifts via the inverse Fourier transform, which
in our formulation only affects $O$ and $\sigma_O$. 

Finally, we separate the model waveform into its five $\ell = 2$ spherical
harmonic modes and calculate the quantities in (\ref{eq:realconst_app})
separately for each mode, which turns $N_1$ and $N_2$ into complex matrices and
$O$ into a vector of discrete inverse Fourier transforms. However, we only need
to calculate those quantities once for a given set of intrinsic binary
parameters and combine them appropriately for each set of orientation and
polarisation angles that we wish to analyse.


\bibliography{paper}

\end{document}

%% file: BinaryFramePrecessing.pdf_tex
\begingroup%
  \makeatletter%
  \providecommand\color[2][]{%
    \errmessage{(Inkscape) Color is used for the text in Inkscape, but the package 'color.sty' is not loaded}%
    \renewcommand\color[2][]{}%
  }%
  \providecommand\transparent[1]{%
    \errmessage{(Inkscape) Transparency is used (non-zero) for the text in Inkscape, but the package 'transparent.sty' is not loaded}%
    \renewcommand\transparent[1]{}%
  }%
  \providecommand\rotatebox[2]{#2}%
  \ifx\svgwidth\undefined%
    \setlength{\unitlength}{339.84539795bp}%
    \ifx\svgscale\undefined%
      \relax%
    \else%
      \setlength{\unitlength}{\unitlength * \real{\svgscale}}%
    \fi%
  \else%
    \setlength{\unitlength}{\svgwidth}%
  \fi%
  \global\let\svgwidth\undefined%
  \global\let\svgscale\undefined%
  \makeatother%
  \begin{picture}(1,0.9257565)%
   
\put(0,0){\includegraphics[width=\unitlength]{BinaryFramePrecessing.pdf}
}%
    \put(0.06328547,0.00622756){\color[rgb]{0,0,0}\makebox(0,0)[lb]{\smash{$x$}}}%
    \put(0.91629006,0.34368655){\color[rgb]{0,0,0}\makebox(0,0)[lb]{\smash{$y$}}}%
    \put(0.37640281,0.89356355){\color[rgb]{0,0,0}\makebox(0,0)[lb]{\smash{$z$}}}%
    \put(0.61465589,0.18523373){\color[rgb]{0,0,0}\makebox(0,0)[lb]{\smash{$m_1$}}}%
    \put(0.03595697,0.50478017){\color[rgb]{0,0,0}\makebox(0,0)[lb]{\smash{$m_2$}}}%
    \put(0.49018854,0.78149908){\color[rgb]{0,0,0}\makebox(0,0)[lb]{\smash{$\vec{S}$}}}%
    \put(0.28649477,0.69629907){\color[rgb]{0,0,0}\makebox(0,0)[lb]{\smash{$\vec{J}_0$}}}%
    \put(0.52853619,0.53037904){\color[rgb]{0,0,0}\makebox(0,0)[lb]{\smash{$\vec{L}$}}}%
    \put(0.38903349,0.24069222){\color[rgb]{0,0,0}\makebox(0,0)[lb]{\smash{$\alpha$}}}%
    \put(0.4201503,0.57745941){\color[rgb]{0,0,0}\makebox(0,0)[lb]{\smash{$\iota$}}}%
    \put(0.27490186,0.44725473){\color[rgb]{0,0,0}\makebox(0,0)[lb]{\smash{$\theta$}}}%
    \put(-0.00331033,0.31957259){\color[rgb]{0,0,0}\makebox(0,0)[lb]{\smash{$\hat{N}$}}}%
  \end{picture}%
\endgroup%